\documentclass[journal]{IEEEtran}
\usepackage{graphicx}
\usepackage[cmex10]{amsmath}
\usepackage{cases}
\usepackage[tight,footnotesize]{subfigure}
\usepackage{amsthm} %proof
\usepackage{cite}
\usepackage{amssymb}
\usepackage{algorithm}
\usepackage{algorithmic}
\usepackage{multirow}
\usepackage{amsmath}
\usepackage{xcolor}
\usepackage{CJK}
\usepackage{subeqnarray}
\usepackage{cases}
\usepackage{enumerate}
\usepackage{cuted}
\usepackage{booktabs}
\newtheorem{theorem}{\hskip\parindent \it Theorem}
\newtheorem{remark}{\hskip\parindent\it{Remark}}

\ifCLASSINFOpdf
\else
\fi

\begin{document}

%\title{ STRA-RIS Enhanced Joint Secure and Covert Communications for Multi-antenna mmWave Systems}%: Analysis and Optimization}
\title{ \LARGE{STAR-RIS Enhanced Joint Physical Layer Security and Covert Communications for Multi-antenna mmWave Systems}}
%\title{ STRA-RIS Enhanced Joint Physical Layer Security and Covert Communications for Multi-antenna mmWave Systems}
%A New Framework for Joint PLS and Covert Communication in Multi-antenna mmWave Systems with STRA-RIS}

\author{Han Xiao,~\IEEEmembership{Student Member,~IEEE,} Xiaoyan Hu$^*$,~\IEEEmembership{Member,~IEEE,}\\
	%Pengcheng Mu,~\IEEEmembership{Member,~IEEE,}  \\ Wenjie Wang,~\IEEEmembership{Member,~IEEE,}  Tong-Xing Zheng,~\IEEEmembership{Member,~IEEE,}
    Ang Li,~\IEEEmembership{Senior Member,~IEEE,} Wenjie Wang,~\IEEEmembership{Senior Member,~IEEE,} \\Zhou Su,~\IEEEmembership{Senior Member,~IEEE,}
    % Christos Masouros,~\IEEEmembership{Senior Member,~IEEE,}
	%Pengcheng Mu,~\IEEEmembership{Member,~IEEE,}  Wenjie Wang,~\IEEEmembership{Member,~IEEE,}\\
	Kai-Kit~Wong,~\IEEEmembership{Fellow,~IEEE}, Kun~Yang,~\IEEEmembership{Fellow,~IEEE}
	
%\thanks{%Manuscript received XX XX, XXXX; revised XX XX, XXXX; accepted XX XX, XXXX.
%This work is supported in part by the National Natural Science Foundation of China (NSFC) under Grant 62201449, and in part by the Key R$\&$D Projects of Shaanixi Province under Grant 2023-YBGY-040, and in part by the Qin Chuang Yuan High-Level Innovation and Entrepreneurship Talent Program under Grant QCYRCXM-2022-231. \emph{(Corresponding author: Xiaoyan Hu.)}}	
	%\thanks{H. Xiao, X. Hu, P. Mu, and W. Wang are with the School of Information and Communication Engineering, Xi'an Jiaotong University, Xi'an 710049, China. (email: hanxiaonuli@stu.xjtu.edu.cn, xiaoyanhu@xjtu.edu.cn, \{pcmu, wjwang\}@mail.xjtu.edu.cn).}
	%\thanks{H. Xiao, X. Hu, P. Mu, W. Wang, and T.-X. Zheng are with the School of Information and Communication Engineering, Xi'an Jiaotong University, Xi'an 710049, China. (email: hanxiaonuli@stu.xjtu.edu.cn, xiaoyanhu@xjtu.edu.cn, \{pcmu, wjwang, zhengtx\}@mail.xjtu.edu.cn).}
	\thanks{H. Xiao, X. Hu,  A. Li, W. Wang, and Z. Su are with the School of Information and Communication Engineering, Xi'an Jiaotong University, Xi'an 710049, China. (email: hanxiaonuli@stu.xjtu.edu.cn, xiaoyanhu@xjtu.edu.cn,  ang.li.2020@xjtu.edu.cn, wjwang@mail.xjtu.edu.cn, zhousu@xjtu.edu.cn).}
	% wjwang@mail.xjtu.edu.cn, zhengtx@mail.xjtu.edu.cn, and the Ministry of Education Key Laboratory for Intelligent Networks and Network Security
    %\thanks{Y. Zhu is with the Department of Electrical and Electronic Engineering, University of Warwick, Coventry CV4 7AL, U.K. (email: yongxu.wong@ucl.ac.uk)}
	\thanks{K.-K. Wong is with the Department of Electronic and Electrical Engineering, University College London, London WC1E 7JE, U.K. (email: kai-kit.wong@ucl.ac.uk)}
	
	\thanks{K. Yang is with the School of Computer Science and Electronic Engineering, University of Essex, Colchester CO4 3SQ, U.K. (e-mail: kunyang@essex.ac.uk).}
}
%$^{\dag}$Department of Electronic and Electrical Engineering, University College London, London, UK\\
%Email:xxxx@xxx

\maketitle

\begin{abstract}
This paper investigates the utilization of simultaneously transmitting and reflecting RIS (STAR-RIS) in supporting joint physical layer security (PLS) and covert communications (CCs) in a multi-antenna millimeter wave (mmWave) system, where the base station (BS) communicates with both covert and security users while defeating eavesdropping by wardens with the help of a STAR-RIS. Specifically, analytical derivations are performed to obtain the closed-form expression of warden's minimum detection error probability (DEP). Furthermore, the asymptotic result of the minimum DEP and the lower bound of the secure rates are derived, considering the practical assumption that BS only knows the statistical channel state information (CSI) between STAR-RIS and the wardens. Subsequently, an optimization problem is formulated with the aim of maximizing the average sum of the covert rate and the minimum secure rate while ensuring the covert requirement and quality of service (QoS) for legal users by jointly optimizing the active and passive beamformers. Due to the strong coupling among variables, an iterative algorithm based on the alternating strategy and the semi-definite relaxation (SDR) method is proposed to solve the non-convex optimization problem. Simulation results indicate that the performance of the proposed STAR-RIS-assisted scheme greatly surpasses that of the conventional RIS scheme, which validates the superiority of STAR-RIS in simultaneously implementing PLS and CCs. %joint PLS and CC
\end{abstract}
\IEEEpeerreviewmaketitle

\vspace{-3mm}
\section{Introduction}\label{sec:S1}
As wireless communication technologies continue to develop rapidly, the security of communications has become a growing concern for both enterprises and individuals.
%This is because open wireless networks often transmit a large amount of sensitive and personal information.
To safeguard users' information from eavesdropping attacks, physical layer security (PLS) has emerged as a promising technique and garnered significant attention in recent years.
As a pioneering work, \cite{Wyner1975wire} demonstrates that a positive perfect secrecy rate can be achieved at the transceiver if the eavesdropper's channel is a diminished form of the legitimate user's channel.
Following this, numerous methods have been proposed with the aim of improving the performance of PLS \cite{yang2012transmit, lv2015secrecy, zhao2017artificial, hu2016secrecy, zheng2022physical}. %zheng2015multi,
In particular, \cite{yang2012transmit} proposes a transmit antenna selection scheme to enhance the PLS considering the practical case without the knowledge of eavesdroppers' channel state information (CSI).
Then \cite{lv2015secrecy} examines the potential of active beamforming to improve the security performance of Heterogeneous networks.
In  \cite{zhao2017artificial}, the utilization of artificial noise (AN) is shown to be beneficial against eavesdropping. % \cite{zheng2015multi} and
The authors of \cite{hu2016secrecy, zheng2022physical} both explore the uncoordinated cooperative jamming schemes to maximize the secure rate while defeating the eavesdropping by appropriately allocating the jamming power.

However, in some scenarios like secret military operations, the security level provided by PLS may not be sufficient.
This is because PLS can only hide the contents of messages but not the existence of communications between authorized users, which may leave security risks that can be exploited by unauthorized users to launch attacks \cite{zheng21}.
Recently, covert communication (CC)  as a novel security technology has drawn great attention from both military and civilian fields\cite{yan2019low}.
CC has the ability to fundamentally conceal the presence of communications between users, providing a higher level of security than PLS.
Toward this end, Bash \textit{et al.} first demonstrate that $\mathcal{O}(\sqrt{n})$ bits of information can be reliably transmitted with a low probability of detection over additive white Gaussian noise (AWGN) channels \cite{bash2013limits}.
Since then, lots of efforts have been made to improve the covert performance \cite{goeckel2015covert, he2017covert, wang2018covert,  chen2021multi, li2020optimal}. %hu2019covert, sobers2017covert, Specifically, \cite{goeckel2015covert} proves that $\mathcal{O}(n)$ bits of covert information can be sent to receivers if the uncertainty of the background noise is considered.
In \cite{he2017covert}, two background noise models with noise uncertainty are proposed based on which the authors investigate the maximum achievable covert rate.
Also, the potential covert performance gain brought by the channel uncertainty is explored in \cite{wang2018covert}.
In \cite{ chen2021multi}, full-duplex receivers are leveraged to transmit jamming signals so as to degrade the detection capabilities of wardens and maximize the covert throughput. %hu2019covert,
The strategy of uninformed jamming is implemented to facilitate CCs between legal users through deliberately generating jamming signals under different channel models \cite{li2020optimal}. %sobers2017covert,
Different from the literature above, \cite{zheng2019multi} investigates the advantages of centralized and distributed multi-antenna transmitters in defeating the wardens with random positions.
Moreover, \cite{shahzad2019covert} exploits the impacts of the number of antennas at the adversary wardens on the covert rate and finds that a slight increase in the number of antennas results in a dramatic decrease in the covert rate.

Although the strategies mentioned above have demonstrated their effectiveness in enhancing the performance of PLS and CCs, it is necessary to acknowledge that their potentials may be highly constrained by the stochastic nature of the wireless propagation environment. %for performance improvement
Specifically, in communication systems operating at millimeter-wave (mmWave) frequencies, this constraint will be particularly pronounced due to the susceptibility of mmWave signals to blockages. % and high propagation losses.
In order to break through this constraint, reconfigurable intelligent surfaces (RISs) emerged as a promising solution % to effectively resist  the randomness of the wireless environment.
%which are usually cost-effective two-dimensional metamaterial surfaces
which consists of numerous cost-effective metamaterial elements and each element can be dynamically controlled so as to modify the electromagnetic characteristics  of the incident signals and reconfigure desirable propagation environments.  %which offers the potential to alter and optimize the propagation environment to meet specific requirements. %(i.e., phases, amplitudes, etc.)
These attractive features of RIS make it popular in both academia and industry, which have been widely investigated in the performance enhancement of wireless applications including PLS and CCs \cite{cui2019secure,  dong2021double, lu2020intelligent, zhou2021intelligent}. %niu2021weighted,%has drawn great attention from both academia and industry which have also been utilized to facilitate the wireless communications applications including PLS and CC \cite{cui2019secure, niu2021weighted, dong2021double, lu2020intelligent, zhou2021intelligent}.
%Hence, this feature attracts a large number of attention from academy and industry and has been utilized to facilitate the wireless communication including PLS and CC \cite{cui2019secure, niu2021weighted, dong2021double, lu2020intelligent, zhou2021intelligent}.
In particular, \cite{cui2019secure} explores the PLS in a RIS-aided multi-antenna communication system with  strong eavesdropping channels, and the achievable secure rate is maximized  by optimizing the active and passive beamformers.
%Niu \textit{et al.} in \cite{niu2021weighted} investigate the PLS performance gain  in a multi-user  system where the AN is leveraged in the RIS-assisted anti-eavesdropping scheme. %multi-input single-output % brought by RIS
%The benefits of utilizing a single reflecting surface (RIS) for improving physical layer security (PLS) have been well-established. However,
To further enhance the secrecy performance, a double RIS scheme incorporating inter-RIS signal reflections is proposed in \cite{dong2021double}.
In addition, \cite{lu2020intelligent} provides a general summary of the potential applications of RIS in enhancing CCs.
The authors in \cite{zhou2021intelligent} examine the performance gain of CCs facilitated by RIS, which indicates that RIS can enable perfect covertness subject to the instantaneous channel state information (CSI) of wardens being accessible.

It is noteworthy that the traditional RIS in the literature above only reflects incident signals, %is narrowly focused on providing a reflective function that
which requires  both transmitters and receivers to situate on the same side of RIS \cite{X.Hu21}.
%Consequently, the utilization of traditional RISs may not offer the required agility and efficiency in cases where users are situated on opposite sides of the RISs.
In order to overcome this limitation, a novel RIS called simultaneously transmitting and reflecting RIS (STAR-RIS) is proposed and developed in \cite{liu2021star, mu2021simultaneously}.
%Specifically, the STAR-RIS separates the incident signals into two parts, where one part is reflected back to the same side of the incident signals while the other part is transmitted to the opposite side.
%For these two parts, the STAR-RIS respectively provides reflected and transmitted coefficients to adjust them, which can construct a full-space smart radio environment with 360$^\circ$ coverage. % to be constructed.
Specifically, the STAR-RIS can separate the incident signals into one reflected part  and one transmitted part which can be dynamically controlled by the reflected and transmitted coefficients so as to construct a full-space smart radio environment with 360$^\circ$ coverage. % to be constructed.
Hence, the STAR-RIS has enormous potential in wireless communications, which has sparked significant interest from both academia and industry \cite{liu2021star}.
However, the research on incorporating STAR-RISs into wireless communication systems is still in its early stages.
In terms of the secure/covert communications, only a small number of works investigate the secure/covert performance gain facilitated by STAR-RIS \cite{han2022artificial, zhang2022secrecy, xiao2023simultaneously,xiao2023star}.
In particular, Han \textit{et al.} \cite{han2022artificial} and Zhang \textit{et al.} \cite{zhang2022secrecy} investigate the potentials of the STAR-RIS in boosting downlink and uplink PLS, respectively. In \cite{xiao2023simultaneously, xiao2023star}, the authors initially explore the potentials of STAR-RIS in CCs, indicating that the STAR-RIS-assisted CCs scheme significantly outperforms the conventional RIS-aided scheme.

In practical scenarios, it is highly possible that users have varying security requirements for communications, e.g., some users may require secure information transmissions and some users may need a higher level of covert communications. In this case, \cite{forouzesh2020joint} first considers a scenario with both PLS and CCs users and analyzes the average sum rate between the secure rate and covert rate under the perfect and imperfect CSI. However, the inherent randomness of the wireless channels results in a limited average rate.
To address this problem, we establish a novel system model enabled by the STAR-RIS for joint implementation of PLS and CCs in this paper. Our main contributions are summarized as follows:

\begin{itemize}
\item \emph{\textbf{STAR-RIS-assisted Joint PLS and CCs Architecture:}}
In this paper, we construct a STAR-RIS-assisted joint PLS and CCs architecture for the first time. This architecture allows the legitimate users who are located on two sides of the STAR-RIS and have varying security needs, e.g., PLS and CCs, to be simultaneously served by elaborately designing the passive reflected and transmitted coefficients of the STAR-RIS as well as the the active transmit beamforming of the base station (BS).
\item \emph{\textbf{Closed-form Expressions of PLS/CCs System Indicators:}}
For CCs, the optimal detection threshold and minimum detection error probability (DEP) at Warden are derived analytically. Additionally, the large system analytic technique is introduced to further derive the asymptotic analytic result of the minimum DEP, which is leveraged as the covert constraint to jointly optimize the active and passive beamformers. For PLS, we derive the close-form expression of a lower bound for the secure rate considering that only the statistical CSI of the eavesdropper is available at the BS. %for the proposed STAR-RIS-aided joint PLS and CC system model , considering the worst case, $\mathbf{h}_\mathrm{re}$
\item \emph{\textbf{Optimization Problem Formulation under Practical Constraints:}}
An optimization problem is formulated for the STAR-RIS-aided joint PLS and CCs system to maximize the average sum rate between the minimum secure rate and covert rate, subject to covert and quality of service (QoS) constraints. This is achieved by jointly optimizing the active and passive beamforming variables. In fact, it is challenging to handle this optimization problem due to the strong coupling among variables especially considering the non-convex amplitude constraint introduced by STAR-RIS.	
\item \emph{\textbf{Iterative Algorithm with Guaranteed Convergence and Substantial Performance Gain:}}
To solve the formulated optimization problem, we propose an iterative algorithm leveraging the alternating strategy and the semi-definite relaxation (SDR) method. Specifically, the optimization problem is divided into two subproblems respectively for active and passive beamforming design. The convergence of the proposed algorithm can be guaranteed which is also verified by simulation results.	
The performance gain of the proposed  STAR-RIS-assisted joint PLS and \mbox{CCs} scheme is demonstrated through comparing with a benchmark scheme utilizing the conventional RIS. The numerical results indicate that STAR-RIS offers greater potential in enhancing the performance of joint PLS and CCs as compared to the traditional RIS. 		
%\item \emph{\textbf{Substantial Performance Gain:}} The performance gain of the proposed  STAR-RIS-assisted joint PLS and \mbox{CCs} scheme is demonstrated through comparing with a benchmark scheme utilizing the conventional RIS. The numerical results indicate that STAR-RIS offers greater potential in enhancing the performance of the joint PLS and CCs as compared to the traditional RIS.  			
\end{itemize}

The remainder of this paper is organized as follows. Section \ref{sec:S2} shows the considered STAR-RIS-aided system model and millimeter wave channel model for joint PLS and CCs. % and establishes the millimeter wave channel model on the basis of the Saleh-Valenzuela channel model.
The minimum DEP and its asymptotic analytic result are analytically derived in Section \ref{sec:S3}. In addition, the close-form expression of the lower bound for the secure rate is also given in Section \ref{sec:S3}. The proposed iterative algorithm and the analysis on its convergence and computational complexity are presented in Section \ref{sec:S4}. Section \ref{sec:S5} gives the numerical simulation results to verify the effectiveness of the proposed scheme. Finally, the conclusion is made in Section \ref{sec:S6}.

\textit{Notation:} Operator $\otimes$ denotes the Kronecker product. $(\cdot)^T$, $(\cdot)^H$ and $(\cdot)^*$ represent transpose, conjugate transpose and conjugate, respectively. $\operatorname{Diag}(\mathbf{a})$ denotes a diagonal matrix with diagonal elements in vector $\mathbf{a}$ while $\operatorname{diag}(\mathbf{A})$ denotes a vector whose elements are composed of the diagonal elements of matrix $\mathbf{A}$. $|\cdot|$, $\|\cdot\|_2$ and $\|\cdot\|_\mathrm{F}$ indicate the complex modulus, the spectral norm and Frobenius norm, respectively. $\mathbb{C}^{M\times N}$ stands for the set of $M\times N$ complex matrices. $x\sim\mathcal{C N}(a,b)$ is the circularly symmetric complex Gaussian random variable with mean $a$ and variance $b$. $\operatorname{Tr}(\cdot)$ represents the trace. $\mathbf{A}\succeq0$ indicates that matrix $\mathbf{A}$ is a positive semidefinite matrix. $\mathbf{I}_{M\times1}$ represents the vector with $M\times 1$ entries that are $1$. $\mathbf{I}_K$ represents the $K\times K$ identity matrix and $\mathbf{e}_k$ is its $k$-th column.

\section{System Model}\label{sec:S2}
\begin{figure}[ht]
	\centering
	\includegraphics[scale=0.4]{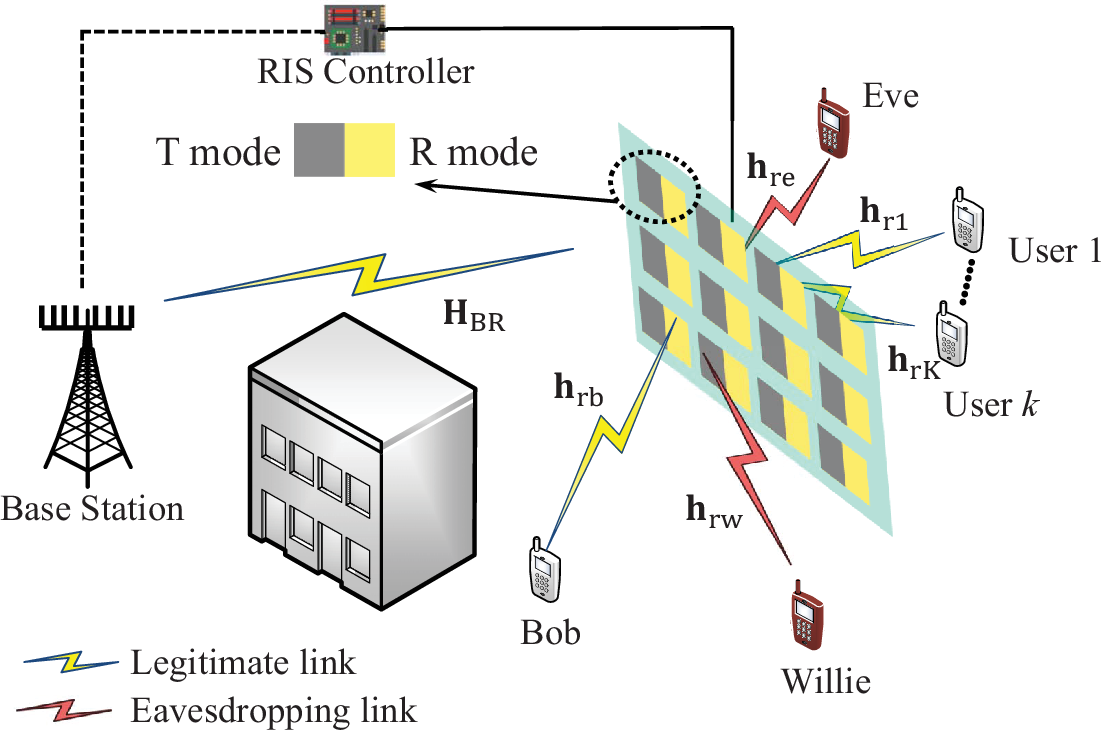}\\
	\caption{System model for STAR-RIS-assisted joint PLS and CC.}\label{fig:System}
\end{figure}

We consider a STAR-RIS-aided system model for joint PLS and CCs as shown in Fig. \ref{fig:System}, which comprises a base station (BS) with $N_\mathrm{t}$ antennas, a covert user (Bob), two eavesdropping/warden users (Willie and Eve), $K$ security users with index $k\in\mathcal{K}\triangleq\{1, 2, \cdots, K\}$, and a STAR-RIS with $M$ elements. All users are equipped with a single antenna and operate in half-duplex mode at the mmWave band.
A practical scenario is investigated where the direct links between Alice and all users are blocked by obstacles such as buildings.
To enhance the communication performance between Alice and legitimate users while impairing the detections by warden users Willie and Eve, an assistant STAR-RIS is deployed near the users.
Without loss of generality, we assume that the covert user Bob and security users are located on the opposite sides of the STAR-RIS, allowing them to be simultaneously served by reflected (R) and transmitted (T) signals, respectively.
The energy splitting protocol is implemented for STAR-RIS whose all elements operate T\&R mode simulataneously\cite{mu2021simultaneously}.\footnote{
Note that the path loss is extremely severe for mmWave communications, we ignore the signals reflected or transmitted more than once by the STAR-RIS in the considered system.}

In this paper, Saleh-Valenzuela channel model \cite{akdeniz2014millimeter} is adopted  for the mmWave communications. In addition, we assume that the uniform linear array (ULA) of antennas is employed at the BS while the STAR-RIS adopts the uniform planar array (UPA). Hence, the channels between BS and STAR-RIS, and between STAR-RIS and users of \{Bob, Willie, Eve, user $k \in \mathcal{K}$\} can be modeled as
\begin{align}
&\hspace{-2mm}\mathbf{H}_\mathrm{BR}=\sqrt{\frac{N_\mathrm{t}M\rho_\mathrm{BR}}{L}}\sum\limits_{l=1}^{L}\varphi^\mathrm{BR}_l\mathbf{a}_\mathrm{R}\left(\phi_l^\mathrm{BR}, \theta_l^\mathrm{BR}\right)\mathbf{a}^H_\mathrm{B}\left(\gamma_l^\mathrm{BR}\right), \label{HBR}\\
&\hspace{-2mm}\mathbf{h}_{\varsigma}=\sqrt{\frac{M\rho_{\varsigma}}{P}}\sum\limits_{p=1}^{P}g^\varsigma_p\mathbf{a}_\mathrm{R}\left(\phi_p^{\varsigma}, \theta_p^{\varsigma}\right)~\mathrm{for}~   \varsigma\in\{\mathrm{rb}, \mathrm{rw}, \mathrm{re}, \mathrm{r}k\}, \label{h_r}
\end{align}
where $\mathbf{H}_\mathrm{BR}\in \mathbb{C}$ and $\mathbf{h}_{\varsigma}\in \mathbb{C}$ with $\rho_\mathrm{BR}$ and $\rho_\mathrm{\varsigma}$ being the path loss values related to BS-RIS link and RIS-users links, respectively.
$L$, $P$ denote the total number of paths in $\mathbf{H}_\mathrm{BR}$ and $\mathbf{h}_\varsigma$, %$N_\mathrm{t}$ and $M$ represent the total number of antennas at BS and elements equipped at STAR-RIS;
and $\varphi_l^\mathrm{BR}$, $g_p^{\varsigma}\sim\mathcal{CN}(0,1)$ are the complex gain of the $l$-th path in $\mathbf{H}_\mathrm{BR}$ and $p$-th path in $\mathbf{h}_\varsigma$, respectively.
%$\zeta_\mathrm{BR}$ and $\zeta_\mathrm{\varsigma}$ for $\varsigma\in\{\mathrm{rb}, \mathrm{rw}, \mathrm{re}, \mathrm{r}k\}$  are the path loss values related to BS-RIS link and RIS-users links for \{Bob, Willie, Eve, user $k$\}, respectively;
Also, $\phi_l^\mathrm{BR}$ and $\theta_l^\mathrm{BR}$ represent the azimuth and elevation angle of arrival (AoA) values at STAR-RIS; $\gamma_l^\mathrm{BR}$ indicates the azimuth angle of departure (AoD) associated with BS; $\phi_p^{\varsigma}$ and $\theta_p^{\varsigma}$ denotes the azimuth and elevation AoD values associated with the  RIS-users links.
In addition, $\mathbf{a}_\mathrm{R}(\phi,\theta)$ and $ \mathbf{a}_\mathrm{B}(\gamma)$ are respectively the beam steering vectors of the ULA and UPA at the  BS and STAR-RIS, which are expressed as
\begin{align}
&\mathbf{a}_\mathrm{B}(\gamma)=\frac{1}{\sqrt{N_\mathrm{t}}}\Big[1, \cdots, e^{j\frac{2\pi d}{\lambda}(n_\mathrm{t}-1)\sin(\gamma)}, \cdots,\notag\\
&\qquad\qquad\qquad~~ e^{j\frac{2\pi d}{\lambda}(N_\mathrm{t}-1)\sin(\gamma) }\Big]^T, \\
& \mathbf{a}_\mathrm{R}(\phi,\theta)\notag\\
&=\frac{1}{\sqrt{M}}\Big[1, \cdots, e^{j\frac{2\pi d}{\lambda}\left(\left(m_\mathrm{y}-1\right)\sin(\phi)\sin(\theta)+(m_\mathrm{z}-1)\cos(\theta)\right)},\notag\\&\cdots, e^{j\frac{2\pi d}{\lambda}\left(\left(M_\mathrm{y}-1\right)\sin(\phi)\sin(\theta)+(M_\mathrm{z}-1)\cos(\theta)\right)}\Big]^T,
\end{align}
where $n_\mathrm{t}\in\{1, 2, \cdots, N_\mathrm{t}\}$, and $m_\mathrm{y}\in\{1, 2, \cdots, M_\mathrm{y}\}$, $m_\mathrm{z}\in\{1, 2, \cdots, M_\mathrm{z}\}$ with $M_\mathrm{y}$, $M_\mathrm{z}$ being the number of elements in horizontal and vertical directions of UPA and $M=M_\mathrm{y}M_\mathrm{z}$.

It is assumed that the considered STAR-RIS-assisted joint PLS and CCs system operates in time division duplex (TDD) mode, enabling the use of STAR-RIS-based uplink channel estimation techniques to obtain the required CSI with channel reciprocity \cite{wu21}. In addition, we assume that the BS has the  knowledge of the instantaneous CSI between STAR-RIS and all legal users, i.e., $\mathbf{H}_\mathrm{BR}$, $\mathbf{h}_{\mathrm{rb}}  $ and $\mathbf{h}_{\mathrm{r}k}$, while  only the statistical CSI between STAR-RIS and Willie/Eve, i.e., $\mathbf{h}_{\mathrm{rw}}$ and $\mathbf{h}_{\mathrm{re}}$, are available at BS. In contrast,  Willie knows the instantaneous CSI of $\mathbf{h}_{\mathrm{rb}}$ and $\mathbf{h}_{\mathrm{r}k}$, but only the statistical CSI of $\mathbf{H}_\mathrm{BR}$ is accessible by Willie, which introduces uncertainty that is beneficial to cover the communications between BS and Bob.
\section{Analysis on the STAR-RIS-Assisted Joint PLS and CCs System }\label{sec:S3}
%\section{Joint PLS and CCs in STAR-RIS-Assisted System }\label{sec:S3}
\subsection{Theoretical Analysis on CCs}
In this section, we focus on the theoretical analysis of the CCs of the system. Specifically, we first discuss Willie's detection strategy for CCs between BS and Bob, and then derive the closed-form expressions of its detection error probability (DEP) and the optimal detection threshold.
Specifically, Willie determines the existence of communications between BS and Bob through the received signal sequences in a time slot, denoted as $\{y_\mathrm{w}[t]\}_{t=1}^T$.
It has to face a binary hypothesis for the judgement of CCs, which includes a null hypothesis $\mathcal{H}_0$, denoting that BS only communicates with the security users without CCs to Bob; and an alternative hypothesis $\mathcal{H}_1$, indicating that there exists CCs between BS and Bob. Under these two hypotheses, the received signals at Bob, Willie and the $k$-th security user can be respectively expressed as
 \begin{align}
	y_\mathrm{b}[t]=&\begin{cases}
		\sum\limits_{k=1}^K\mathbf{h}_\mathrm{rb}^H\boldsymbol{\Theta}_\mathrm{r}\mathbf{H}_\mathrm{BR}\mathbf{w}_ks_k[t]+n_\mathrm{b}[t], & \mathcal{H}_0,\\
		\mathbf{h}_\mathrm{rb}^H\boldsymbol{\Theta}_\mathrm{r}\mathbf{H}_\mathrm{BR}\mathbf{w}_\mathrm{b}s_\mathrm{b}[t]+\\  \sum\limits_{k=1}^K\mathbf{h}_\mathrm{rb}^H\boldsymbol{\Theta}_\mathrm{r}\mathbf{H}_\mathrm{BR}\mathbf{w}_ks_k[t]+n_\mathrm{b}[t], &\mathcal{H}_1,
	\end{cases}\\
    	y_\mathrm{w}[t]=&\begin{cases}
    		\sum\limits_{k=1}^K\mathbf{h}_\mathrm{rw}^H\boldsymbol{\Theta}_\mathrm{r}\mathbf{H}_\mathrm{BR}\mathbf{w}_ks_k[t]+n_\mathrm{w}[t], & \mathcal{H}_0,\\
    		\mathbf{h}_\mathrm{rw}^H\boldsymbol{\Theta}_\mathrm{r}\mathbf{H}_\mathrm{BR}\mathbf{w}_\mathrm{b}s_\mathrm{b}[t]+\\  \sum\limits_{k=1}^K\mathbf{h}_\mathrm{rw}^H\boldsymbol{\Theta}_\mathrm{r}\mathbf{H}_\mathrm{BR}\mathbf{w}_ks_k[t]+n_\mathrm{w}[t], &\mathcal{H}_1,
    	\end{cases}\\
    	y_k=&\begin{cases}
    	\sum\limits_{k=1}^K\mathbf{h}_{\mathrm{r}k}^H\boldsymbol{\Theta}_\mathrm{t}\mathbf{H}_\mathrm{BR}\mathbf{w}_ks_k[t]+n_k[t], & \mathcal{H}_0,\\
    	\mathbf{h}_{\mathrm{r}k}^H\boldsymbol{\Theta}_\mathrm{t}\mathbf{H}_\mathrm{BR}\mathbf{w}_\mathrm{b}s_\mathrm{b}[t]+\\  \sum\limits_{k=1}^K\mathbf{h}_{\mathrm{r}k}^H\boldsymbol{\Theta}_\mathrm{t}\mathbf{H}_\mathrm{BR}\mathbf{w}_ks_k[t]+n_k[t], &\mathcal{H}_1,
    \end{cases}\label{eq_receive_k}
\end{align}
where $t \in \mathcal{T}\triangleq\{1, \cdots, T\}$ is the index of each communication channel use with the maximum number of $T$ in a time slot.
$\boldsymbol{\Theta}_\xi=\operatorname{Diag}\Big\{\sqrt{\beta_\xi^1} e^{\mathrm{j} \phi_\xi^1}, \ldots, \sqrt{\beta_\xi^M}e^{\mathrm{j}\phi_\xi^M}\Big\}$ indicates the reflected or transmitted coefficient matrix of STAR-RIS with $\xi\in\{\mathrm{r}, \mathrm{t}\}$, where $\beta_\xi^m \in[0,1]$, $\phi_\xi^m\in[0,2\pi)$ and $\beta_\mathrm{r}^m+\beta_\mathrm{t}^m=1$, for $\forall m \in \mathcal{M} \triangleq\{1,2, \ldots, M\}$.
Also, $s_\mathrm{b}$, $s_k\sim\mathcal{C N}(0, 1)$ respectively represent the signals transmitted by BS to Bob and the $k$-th security user,
while $\mathbf{w}_\mathrm{b}$, $\mathbf{w}_k\in\mathbb{C}^{N_\mathrm{t}\times1}$ are the beamforming vectors correspondingly.
In addition, $n_\mathrm{b}\sim\mathcal{CN}(0, \sigma^2_{\mathrm{b}})$, $n_\mathrm{w}\sim\mathcal{CN}(0, \sigma^2_{\mathrm{w}})$ and $n_{k}\sim\mathcal{CN}(0, \sigma^2_k)$ for $k\in\mathcal{K}$ denote the AWGN noise at Bob, Willie and the $k$-th security user.

We assume that Willie leverages a radiometer to detect CCs between BS and Bob, %. In accordance with the radiometer's working mechanism, Willie uses
where the average power of the received signals, i.e., $\overline{P}_\mathrm{w}=\frac{1}{T}\sum_{t=1}^{T}\left|y_\mathrm{w}[t]\right|^2$, is used to do the statistical test. In line with existing works (e.g., \cite{xiao2023simultaneously, zheng21}), it is assumed that Willie utilizes an infinite number of signal samples, i.e., $T\rightarrow \infty$, to judge the binary hypotheses. Hence, the received average power can be derived as
\begin{align}
	\overline{P}_\mathrm{w}=&\lim_{T\rightarrow \infty}\frac{1}{T}\sum_{t=1}^{T}\left|y_\mathrm{w}[t]\right|^2\notag\\
	=&\begin{cases}
		\sum\limits_{k=1}^K\left|\mathbf{h}_\mathrm{rw}^H\boldsymbol{\Theta}_\mathrm{r}\mathbf{H}_\mathrm{BR}\mathbf{w}_k|\right.^2+\sigma^2_\mathrm{w}, &\mathcal{H}_0,\\
		\left|\mathbf{h}_\mathrm{rw}^H\boldsymbol{\Theta}_\mathrm{r}\mathbf{H}_\mathrm{BR}\mathbf{w}_\mathrm{b}|\right.^2+\\
		\sum\limits_{k=1}^K\left|\mathbf{h}_\mathrm{rw}^H\boldsymbol{\Theta}_\mathrm{r}\mathbf{H}_\mathrm{BR}\mathbf{w}_k|\right.^2+\sigma^2_\mathrm{w}, &\mathcal{H}_1.
	\end{cases}
\end{align}

To determine the existence of CCs between BS and Bob, Willie needs to analyze $\overline{P}_\mathrm{w}$ under the hypotheses of $\mathcal{H}_0$ and $\mathcal{H}_1$  by leveraging the decision rule $\overline{P}_\mathrm{w} \underset{\mathcal{D}_0}{\stackrel{\mathcal{D}_1}{\gtrless}} \tau_\mathrm{dt}$, where $\mathcal{D}_0$ (or $\mathcal{D}_1$) represents the decision that Willie favors $\mathcal{H}_0$ (or $\mathcal{H}_1$) and  $\tau_\mathrm{dt}$ is the corresponding detection threshold.
In this paper, we adopt DEP to characterize Willie's detection ability for CCs between BS and Bob, considering the worst-case scenario where Willie can optimize $\tau_\mathrm{dt}$ to obtain the optimal detection threshold and the minimum DEP.
Next, we will analytically derive the minimum DEP based on the false alarm (FA) probability and the miss detection (MD) probability from Willie's perspective, where FA indicates that Willie makes decision $\mathcal{D}_1$ under hypothesis $\mathcal{H}_0$ with probability $P_\mathrm{FA}=\operatorname{Pr}(\mathcal{D}_1|\mathcal{H}_0)$ while MD represents that Willie makes decision $\mathcal{D}_0$ under hypothesis $\mathcal{H}_1$ with probability $P_\mathrm{MD}=\operatorname{Pr}(\mathcal{D}_0|\mathcal{H}_1)$. Specifically, $P_\mathrm{FA}$ and $P_\mathrm{MD}$ are given by Theorem \ref{eq_theorem1}.
\begin{theorem}\label{eq_theorem1}
	The FA probability and MD probability are expressed as
\begin{align}
	P_\mathrm{FA}=&\begin{cases}1,&\tau_\mathrm{dt}\leq\sigma_\mathrm{w}^2,\\
		e^{-\frac{\tau_\mathrm{dt}-\sigma_\mathrm{w}^2}{\lambda_0}},&\mathrm{otherwise},
	\end{cases}\label{eq_PFA}\\
P_\mathrm{MD} =&\begin{cases}0,&\tau_\mathrm{dt}\leq\sigma_\mathrm{w}^2,\\
	1-e^{-\frac{\tau_\mathrm{dt}-\sigma_\mathrm{w}^2}{\lambda_1}},&\mathrm{otherwise},
	\end{cases}\label{eq_PMD}
\end{align}	
where
\begin{itemize}
\item\emph{$\lambda_0=\frac{N_\mathrm{t}M\rho_\mathrm{BR}}{L}\sum\limits_{k=1}^K\left\|\boldsymbol{\Phi}\operatorname{vec}\left((\mathbf{w}_k\mathbf{h}_\mathrm{rw}^H\boldsymbol{\Theta}_\mathrm{r})^T\right)\right\|^2_2$},
\item\emph{$\lambda_1=\frac{N_\mathrm{t}M\rho_\mathrm{BR}}{L}\left\|\boldsymbol{\Phi}\operatorname{vec}\left((\mathbf{w}_\mathrm{b}\mathbf{h}_\mathrm{rw}^H\boldsymbol{\Theta}_\mathrm{r})^T\right)\right\|^2_2+\lambda_0$},
	\item\emph{$\boldsymbol{\Phi}=\left[\operatorname{vec}(\mathbf{A}_1), \cdots,\operatorname{vec}( \mathbf{A}_{L})\right]^H$},
	\item\emph{$\mathbf{A}_{l}=\mathbf{a}_\mathrm{R}\left(\phi_l^\mathrm{BR},\theta_l^\mathrm{BR}\right)\mathbf{a}_\mathrm{B}^H\left(\gamma_l^\mathrm{BR}\right)$}.
\end{itemize}
\begin{proof}
   The proof is given in Appendix \ref{Apd_A}.
\end{proof}
\end{theorem}
According to the piecewise functions in \eqref{eq_PFA} and \eqref{eq_PMD}, we can find that $\sigma_\mathrm{w}^2$ is an important boundary that affects the value of $P_\mathrm{FA}$ and $P_\mathrm{MD}$. Specifically, when $\tau_\mathrm{dt}\leq\sigma_\mathrm{w}^2$, FA will always be performed, but MD can be completely avoided. And with the increase of $\tau_\mathrm{dt}$ from 0 to $\infty$, $P_\mathrm{FA}$ will experience a decrease from 1 to 0, while $P_\mathrm{MD}$ has an opposite trend. Based on the analytical expression of $P_\mathrm{FA}$ and $P_\mathrm{MD}$ in \eqref{eq_PFA} and \eqref{eq_PMD}, Willie's DEP can be derived as
\vspace{-2mm}\begin{align}\label{eq_Pe}
	P_\mathrm{e}=&P_\mathrm{FA}+	P_\mathrm{MD}\notag\\
	=&\begin{cases}
	1, &\tau_\mathrm{dt}\leq \sigma_\mathrm{w}^2,\\
	1-e^{-\frac{\tau_\mathrm{dt}-\sigma_\mathrm{w}^2}{\lambda_1}}+e^{-\frac{\tau_\mathrm{dt}-\sigma_\mathrm{w}^2}{\lambda_0}}, &\mathrm{otherwise}.
	\end{cases}
\end{align}
In this paper, we focus on the uncertain scenario with detection threshold $\tau_\mathrm{dt}>\sigma_\mathrm{w}^2$. Next, we will analyze and derive the optimal detection threshold, denoted as $\tau_\mathrm{dt}^*$, and the minimum DEP $P_\mathrm{e}^*$. In particular, the first-order partial derivative of $P_\mathrm{e}$ with respect to (w.r.t.) $\tau_\mathrm{dt}$ is given by
\vspace{-2mm}\begin{align} \frac{\partial{P_\mathrm{e}}}{\partial{\tau_\mathrm{dt}}}=\frac{e^{-\frac{\tau_\mathrm{dt}-\sigma_\mathrm{w}^2}{\lambda_1}}}{\lambda_1}-\frac{e^{-\frac{\tau_\mathrm{dt}-\sigma_\mathrm{w}^2}{\lambda_0}}}{\lambda_0}.
\end{align}
Let $\frac{\partial{P_\mathrm{e}}}{\partial{\tau_\mathrm{dt}}}=0$, we can obtain the unique solution of $\tau_\mathrm{dt}=\frac{\lambda_1\lambda_0\ln{\frac{\lambda_1}{\lambda_0}}}{\lambda_1-\lambda_0}+\sigma_\mathrm{w}^2$. It is easy to verify that $\frac{\partial{P_\mathrm{e}}}{\partial{\tau_\mathrm{dt}}}>0$ when $\tau_\mathrm{dt}>\frac{\lambda_1\lambda_0\ln{\frac{\lambda_1}{\lambda_0}}}{\lambda_1-\lambda_0}+\sigma_\mathrm{w}^2$, while $\frac{\partial{P_\mathrm{e}}}{\partial{\tau_\mathrm{dt}}}<0$ when $\tau_\mathrm{dt}<\frac{\lambda_1\lambda_0\ln{\frac{\lambda_1}{\lambda_0}}}{\lambda_1-\lambda_0}+\sigma_\mathrm{w}^2$.
Hence, the optimal detection threshold minimizing $P_\mathrm{e}$ can be expressed as $\tau_\mathrm{dt}^*=\frac{\lambda_1\lambda_0\ln{\frac{\lambda_1}{\lambda_0}}}{\lambda_1-\lambda_0}+\sigma_\mathrm{w}^2$ and the corresponding minimum DEP is derived as
\begin{align}\label{eq_minimum_DEP}
	P_\mathrm{e}^*=1-e^{-\frac{\lambda_0\ln{\frac{\lambda_1}{\lambda_0}}}{\lambda_1-\lambda_0}}+e^{-\frac{\lambda_1\ln{\frac{\lambda_1}{\lambda_0}}}{\lambda_1-\lambda_0}}.
\end{align}

In order to guarantee the covertness of communications between BS and Bob, 	$P_\mathrm{e}^*\geq 1-\epsilon $ is required where $\epsilon\in(0,1)$ is a
quite small value required by the system performance indicators. Considering that only the statistical CSI of $\mathbf{h}_\mathrm{rw}$ is available at BS, the average minimum DEP over $\mathbf{h}_\mathrm{rw}$, i.e., $\overline{P}_\mathrm{e}^*=\mathbb{E}_{\mathbf{h}_\mathrm{rw}}(P_\mathrm{e}^*)$, is utilized to evaluate the covert performance.
However, in \eqref{eq_minimum_DEP}, $\lambda_0$ and $\lambda_1$ are both random functions of $\mathbf{h}_\mathrm{rw}$ and are coupled with each other, which makes it challenging to directly calculate $\overline{P}_\mathrm{e}^*$.
To tackle this problem, the large system analytic technique is leveraged to handle the coupling between $\lambda_0$ and $\lambda_1$, which is widely adopted to analyze the performance limitations of wireless communication systems (e.g., \cite{xiao2023simultaneously, Wang21}).
Specifically, we assume that a large number of low-cost elements are equipped at STAR-RIS, and thus the asymptotic analytic results of $\lambda_0$ and $\lambda_1$ can be obtained as in the following Theorem \ref{eq_theorem2}.
\begin{theorem}\label{eq_theorem2}
	Applying the large system analytic technique on $\lambda_0$ and $\lambda_1$, the asymptotic analytic results are given by
\begin{align}	\widehat{\lambda}_0&=\frac{N_\mathrm{t}M^2\rho_\mathrm{BR}\rho_\mathrm{rw}}{LP}\sum\limits_{k=1}^K\sum\limits_{l=1}^{L}\left(\mathbf{w}^H_k\boldsymbol{\Psi}^l_\mathrm{BR}\mathbf{w}_k\right)\notag\\&	\left(\boldsymbol{\vartheta}_\mathrm{r}^T\boldsymbol{\Xi}^T\left(\left(\widehat{\boldsymbol{\Psi}}^l_\mathrm{BR}\right)^T\otimes\left(\boldsymbol{\Omega}_\mathrm{rw}^H\boldsymbol{\Omega}_\mathrm{rw}\right)\right)\boldsymbol{\Xi}\boldsymbol{\vartheta}_\mathrm{r}^*\right),\label{eq_lambda_0a}\\	 	\widehat{\lambda}_1&=\widehat{\lambda}_0+\frac{N_\mathrm{t}M^2\rho_\mathrm{BR}\rho_\mathrm{rw}}{LP}\sum\limits_{l=1}^{L}\left(\mathbf{w}^H_\mathrm{b}\boldsymbol{\Psi}^l_\mathrm{BR}\mathbf{w}_\mathrm{b}\right)\notag \\&	\left(\boldsymbol{\vartheta}_\mathrm{r}^T\boldsymbol{\Xi}^T\left(\left(\widehat{\boldsymbol{\Psi}}^l_\mathrm{BR}\right)^T\otimes\left(\boldsymbol{\Omega}_\mathrm{rw}^H\boldsymbol{\Omega}_\mathrm{rw}\right)\right)\boldsymbol{\Xi}\boldsymbol{\vartheta}_\mathrm{r}^*\right),\label{eq_lambda_1a}
\end{align}
where
\begin{itemize}
	\item\emph{$\boldsymbol{\vartheta}_\mathrm{r}=\operatorname{diag}(\boldsymbol{\Theta}_\mathrm{r}),~
		\boldsymbol{\Psi}^l_\mathrm{BR}=\mathbf{a}_\mathrm{B}\left(\gamma_l^{\mathrm{BR}}\right)\mathbf{a}_\mathrm{B}^H\left(\gamma_l^{\mathrm{BR}}\right)$} ,
	\item\emph{$\widehat{\boldsymbol{\Psi}}^l_\mathrm{BR}=\mathbf{a}_\mathrm{R}\left(\phi_l^{\mathrm{BR}}, \theta_l^{\mathrm{BR}}\right)\mathbf{a}_\mathrm{R}^H\left(\phi_l^\mathrm{BR}, \theta_l^{\mathrm{BR}}\right)$},
	\item\emph{$\boldsymbol{\Omega}_\mathrm{rw}=\left[\mathbf{a}_\mathrm{R}\left(\phi_1^\mathrm{rw}, \theta_1^\mathrm{rw}\right), \cdots,
	\mathbf{a}_\mathrm{R}\left(\phi_{P}^\mathrm{rw}, \theta_{P}^\mathrm{rw}\right)\right]^H$},
\item\emph{$\boldsymbol{\Xi}=\Big[[\mathbf{e}_{1}, \mathbf{0}_{M\times(M-1)}];~[\mathbf{0}_{M\times1}, \mathbf{e}_{2}, \mathbf{0}_{M\times(M-2)}];~\cdots;$ $ [\mathbf{0}_{M\times(M-1)}, \mathbf{e}_{M} ]\Big]$}.
\end{itemize}	
\begin{proof}
	The proof is given in Appendix \ref{Apd_B}.
\end{proof}
\end{theorem}

Thus, we can further obtain the asymptotic analytic result of the minimum DEP by substituting \eqref{eq_lambda_0a} and \eqref{eq_lambda_1a} into \eqref{eq_minimum_DEP} and adopting some algebraic manipulations, which is expressed as
\begin{align}\label{eq_pea}
	  P_\mathrm{ea}^*=&1-e^{-\frac{\beta\ln{\frac{\alpha+\beta}{\beta}}}{\alpha}}\left(1-\frac{\beta}{\alpha+\beta}\right),
\end{align}
where
\begin{itemize}		\item\emph{$\alpha=\frac{N_\mathrm{t}M^2\rho_\mathrm{BR}\rho_\mathrm{rw}}{LP}\sum\limits_{l=1}^{L}\big(\mathbf{w}^H_\mathrm{b}\boldsymbol{\Psi}^l_\mathrm{BR}\mathbf{w}_\mathrm{b}\big)
\Big(\boldsymbol{\vartheta}_\mathrm{r}^T\boldsymbol{\Xi}^T\Big(\big(\widehat{\boldsymbol{\Psi}}^l_\mathrm{BR}\big)^T$ $\otimes\left(\boldsymbol{\Omega}_\mathrm{rw}^H\boldsymbol{\Omega}_\mathrm{rw}\right)\Big)\boldsymbol{\Xi}\boldsymbol{\vartheta}_\mathrm{r}^*\Big)$},		\item\emph{$\beta=\frac{N_\mathrm{t}M^2\rho_\mathrm{BR}\rho_\mathrm{rw}}{LP}\sum\limits_{k=1}^K\sum\limits_{l=1}^{L}\big(\mathbf{w}^H_k\boldsymbol{\Psi}^l_\mathrm{BR}\mathbf{w}_k\big)
	\Big(\boldsymbol{\vartheta}_\mathrm{r}^T\boldsymbol{\Xi}^T\Big(\big(\widehat{\boldsymbol{\Psi}}^l_\mathrm{BR}\big)^T$ $\otimes\left(\boldsymbol{\Omega}_\mathrm{rw}^H\boldsymbol{\Omega}_\mathrm{rw}\right)\Big)\boldsymbol{\Xi}\boldsymbol{\vartheta}_\mathrm{r}^*\Big)$}.
\end{itemize}
In the following sections, the convert constraint $P_\mathrm{ea}^*\geq 1-\epsilon $ will be utilized to characterize and guarantee the covert performance of the system.
\begin{remark}\label{remark1}
Based on the \eqref{eq_pea}, several key conclusions can be made. Specifically, the first-order partial derivative of $P_\mathrm{ea}^*$ w.r.t. $\alpha$ can be expressed as
 \begin{align}
 	\frac{\partial{P_\mathrm{ea}^*}}{\partial{\alpha}}=-\frac{\left(\frac{\alpha+\beta}{\beta}\right)^{-\frac{\alpha+\beta}{\alpha}}\ln\left(\frac{\alpha+\beta}{\beta}\right)}{\alpha}<0,
 \end{align}
which shows that $P_\mathrm{ea}^*$ is a monotonically decreasing function regarding $\alpha$. It is evident that as $\alpha$ increases, Willie's detection ability will improve. Furthermore, by utilizing the L'Hospital's rule, we can determine the limit value of $P_\mathrm{ea}^*$ as $\alpha\rightarrow +\infty$, i.e., $\lim_{\alpha\rightarrow+\infty}P_\mathrm{ea}^*=0$, which is consistent with this fact that as the transmit power allocated to Bob increases, the posssibility of Willie detecting the communications between Alice and Bob also increases.
  \begin{align}\label{eq_partial_beta}
 	\frac{\partial{P_\mathrm{ea}^*}}{\partial{\beta}}=\frac{\left(\frac{\alpha+\beta}{\beta}\right)^{-\frac{\beta}{\alpha}}\ln\left(\frac{\alpha+\beta}{\beta}\right)}{\alpha+\beta}>0.
 \end{align}
Moreover, we can derive the first-order partial derivative of $P_\mathrm{ea}^*$ w.r.t. $\beta$, revealing that $P_\mathrm{ea}^*$ exhibits a monotonically increasing behavior in relation to $\beta$, as shown in equation \eqref{eq_partial_beta}. Moreover, by applying the L'Hospital's rule, it can be determined that $\lim_{\beta\rightarrow+\infty}P_\mathrm{ea}^*=1$, indicating that the security users' signals  have the potential to degrade the detection performance at Willie.
\end{remark}

Note that, when hypothesis $\mathcal{H}_1$ is true, the available covert rate at Bob can be expressed as
\begin{align}
	R^\mathrm{c}_\mathrm{b}=\log_2\left(1+\frac{\left|\mathbf{h}^H_{\mathrm{rb}}\boldsymbol{\Theta}_\mathrm{r}\mathbf{H}_\mathrm{BR}\mathbf{w}_\mathrm{b}\right|^2}{\sum\limits_{k=1}^K\left|\mathbf{h}^H_{\mathrm{rb}}\boldsymbol{\Theta}_\mathrm{r}\mathbf{H}_\mathrm{BR}\mathbf{w}_k\right|^2+\sigma_\mathrm{b}^2}\right).
\end{align}
\subsection{Theoretical Analysis on PLS}
In this section, the theoretical analysis on the PLS of the system is addressed, where we analytically derive the secure rate of all the security users considering that BS only knows the statistical CSI of $\mathbf{h}_\mathrm{re}$. In fact, the secure performance of security users is highly affected by the covert hypotheses. Specifically, the signals received by the $k$-th security user are given by equation \eqref{eq_receive_k}, while the signals received by Eve can be expressed as.
\begin{align}
y_\mathrm{e}=&\begin{cases}
	\sum\limits_{k=1}^K\mathbf{h}_\mathrm{re}^H\boldsymbol{\Theta}_\mathrm{t}\mathbf{H}_\mathrm{BR}\mathbf{w}_ks_k+n_\mathrm{e}, & \mathcal{H}_0,\\
	\mathbf{h}_\mathrm{re}^H\boldsymbol{\Theta}_\mathrm{t}\mathbf{H}_\mathrm{BR}\mathbf{w}_\mathrm{b}s_\mathrm{b}+\\  \sum\limits_{k=1}^K\mathbf{h}_\mathrm{re}^H\boldsymbol{\Theta}_\mathrm{t}\mathbf{H}_\mathrm{BR}\mathbf{w}_ks_k+n_\mathrm{e}, &\mathcal{H}_1.
\end{cases}
\end{align}
%Therefore, the eavesdropping rate at Eve towards the $k$-th security user can be expressed as
%\begin{align}
%	R_\mathrm{e}^k=\begin{cases}
%		\log_2\left(1+\gamma_{e0}\right), & \mathcal{H}_0,\\
%			\log_2\left(1+\gamma_{e1}\right), & \mathcal{H}_1,
%	\end{cases}
%\end{align}
%where $\gamma_{e1}=\frac{\left|\mathbf{h}^H_{\mathrm{re}}\boldsymbol{\Theta}_\mathrm{t}\mathbf{H}_\mathrm{BR}\mathbf{w}_k\right|^2}
%{\left|\mathbf{h}^H_{\mathrm{re}}\boldsymbol{\Theta}_\mathrm{t}\mathbf{H}_\mathrm{BR}\mathbf{w}_\mathrm{b}\right|^2+\sum\limits_{j\neq k}^K\left|\mathbf{h}^H_{\mathrm{re}}\boldsymbol{\Theta}_\mathrm{t}\mathbf{H}_\mathrm{BR}\mathbf{w}_j\right|^2+\sigma_\mathrm{e}^2}$ and $\gamma_{e0}=\frac{\left|\mathbf{h}^H_{\mathrm{re}}\boldsymbol{\Theta}_\mathrm{t}\mathbf{H}_\mathrm{BR}\mathbf{w}_k\right|^2}
%{\sum\limits_{j\neq k}^K\left|\mathbf{h}^H_{\mathrm{re}}\boldsymbol{\Theta}_\mathrm{t}\mathbf{H}_\mathrm{BR}\mathbf{w}_j\right|^2+\sigma_\mathrm{e}^2}$.

Therefore, the secure rate for the $k$-th user is given as
\begin{align}
	R^k_\mathrm{s}=\begin{cases}\left[\log_2\left(1+\gamma_{k0}\right)-\log_2\left(1+\gamma_{\mathrm{e0}}\right)\right]^+, &\mathcal{H}_0,\\
		\left[\log_2\left(1+\gamma_{k1}\right)-\log_2\left(1+\gamma_{\mathrm{e1}}\right)\right]^+, &\mathcal{H}_1,
	\end{cases}
\end{align}
where $\gamma_{k1}=\frac{\left|\mathbf{h}^H_{\mathrm{r}k}\boldsymbol{\Theta}_\mathrm{t}\mathbf{H}_\mathrm{BR}\mathbf{w}_k\right|^2}
{\left|\mathbf{h}^H_{\mathrm{r}k}\boldsymbol{\Theta}_\mathrm{t}\mathbf{H}_\mathrm{BR}\mathbf{w}_\mathrm{b}\right|^2+\sum\limits_{j\neq k}^K\left|\mathbf{h}^H_{\mathrm{r}k}\boldsymbol{\Theta}_\mathrm{t}\mathbf{H}_\mathrm{BR}\mathbf{w}_j\right|^2+\sigma^2_{k}}$,
$\gamma_{k0}=\frac{\left|\mathbf{h}^H_{\mathrm{r}k}\boldsymbol{\Theta}_\mathrm{t}\mathbf{H}_\mathrm{BR}\mathbf{w}_k\right|^2}
	{\sum\limits_{j\neq k}^K\left|\mathbf{h}^H_{\mathrm{r}k}\boldsymbol{\Theta}_\mathrm{t}\mathbf{H}_\mathrm{BR}\mathbf{w}_j\right|^2+\sigma^2_{k}}$, $\gamma_{e0}=\frac{\left|\mathbf{h}^H_{\mathrm{re}}\boldsymbol{\Theta}_\mathrm{t}\mathbf{H}_\mathrm{BR}\mathbf{w}_k\right|^2}
	{\sum\limits_{j\neq k}^K\left|\mathbf{h}^H_{\mathrm{re}}\boldsymbol{\Theta}_\mathrm{t}\mathbf{H}_\mathrm{BR}\mathbf{w}_j\right|^2+\sigma_\mathrm{e}^2}$, $\gamma_{e1}=\frac{\left|\mathbf{h}^H_{\mathrm{re}}\boldsymbol{\Theta}_\mathrm{t}\mathbf{H}_\mathrm{BR}\mathbf{w}_k\right|^2}
	{\left|\mathbf{h}^H_{\mathrm{re}}\boldsymbol{\Theta}_\mathrm{t}\mathbf{H}_\mathrm{BR}\mathbf{w}_\mathrm{b}\right|^2+\sum\limits_{j\neq k}^K\left|\mathbf{h}^H_{\mathrm{re}}\boldsymbol{\Theta}_\mathrm{t}\mathbf{H}_\mathrm{BR}\mathbf{w}_j\right|^2+\sigma_\mathrm{e}^2}$, for $k\in \mathcal{K}$.
Due to the fact that the BS can only acquire the statistical CSI of $\mathbf{h}_\mathrm{re}$, the average secure rates over $\mathbf{h}_\mathrm{re}$
are leveraged. In detail, the average secure rate for the $k$-th user can be further expressed as % to characterize \textcolor[rgb]{1.00,0.00,0.00}{the secure rates in next sections}
\begin{align}
	&\widehat{R}^k_\mathrm{s}=\notag\\&\begin{cases}\left[\log_2\left(1+\gamma_{k0}\right)-\mathbb{E}_{\mathbf{h}_\mathrm{re}}\left(\log_2\left(1+\gamma_{\mathrm{e0}}\right)\right)\right]^+, &\mathcal{H}_0,\\
		\left[\log_2\left(1+\gamma_{k1}\right)-\mathbb{E}_{\mathbf{h}_\mathrm{re}}\left(\log_2\left(1+\gamma_{\mathrm{e1}}\right)\right)\right]^+,& \mathcal{H}_1.
	\end{cases}
\end{align}
It is easy to verify that $\mathbf{h}_\mathrm{re}\sim\mathcal{C N}\left(\mathbf{0}_{M \times1},\frac{M\rho_\mathrm{re}}{P}\boldsymbol{\Omega}_\mathrm{re}^H\boldsymbol{\Omega}_\mathrm{re}\right)$, where $\boldsymbol{\Omega}_\mathrm{re}=\left[\mathbf{a}_\mathrm{R}\left(\phi_1^\mathrm{re}, \theta_1^\mathrm{re}\right), \cdots,
\mathbf{a}_\mathrm{R}\left(\phi_{P}^\mathrm{re}, \theta_{P}^\mathrm{re}\right)\right]^H$.
Hence, the average eavesdropping rates, denoted as $\mathbb{E}_{\mathbf{h}_\mathrm{re}}\left(\log_2\left(1+\gamma_{\mathrm{e0}}\right)\right)$ and $\mathbb{E}_{\mathbf{h}_\mathrm{re}}\left(\log_2\left(1+\gamma_{\mathrm{e1}}\right)\right)$, can be derived as
 \begin{align}
	 &\mathbb{E}_{\mathbf{h}_\mathrm{re}}\left(\log_2\left(1+\gamma_{\mathrm{e0}}\right)\right)	=\frac{e^{\frac{\sigma^2_\mathrm{e}}{\eta_0}}\Gamma(0, \frac{\sigma^2_\mathrm{e}}{\eta_0})}{\ln2}-\frac{e^{\frac{\sigma^2_\mathrm{e}}{\widehat{\eta}_0}}\Gamma(0, \frac{\sigma^2_\mathrm{e}}{\widehat{\eta}_0})}{\ln2},\\
&\mathbb{E}_{\mathbf{h}_\mathrm{re}}\left(\log_2\left(1+\gamma_{\mathrm{e1}}\right)\right)=\frac{e^{\frac{\sigma^2_\mathrm{e}}{\eta_1}}\Gamma(0, \frac{\sigma^2_\mathrm{e}}{\eta_1})}{\ln2}-\frac{e^{\frac{\sigma^2_\mathrm{e}}{\widehat{\eta}_1}}\Gamma(0, \frac{\sigma^2_\mathrm{e}}{\widehat{\eta}_1})}{\ln2},
\end{align}
where $\Gamma(\cdot, \cdot)$ is the upper incomplete Gamma function, and
\begin{itemize}
	\item\emph{$\eta_0=\frac{M\rho_\mathrm{re}}{P}\sum\limits_{k=1}^K\left\|\boldsymbol{\Omega} _\mathrm{re} \boldsymbol{\Theta}_\mathrm{t}\mathbf{H}_\mathrm{BR}\mathbf{w}_k\right\|^2$},
	\item\emph{$	\widehat{\eta}_0=\frac{M\rho_\mathrm{re}}{P}\sum\limits_{j\neq k}^K\left\|\boldsymbol{\Omega}_\mathrm{re} \boldsymbol{\Theta}_\mathrm{t}\mathbf{H}_\mathrm{BR}\mathbf{w}_j\right\|^2$},
	\item\emph{$	\eta_1=\frac{M\rho_\mathrm{re}}{P}\left\|\boldsymbol{\Omega}_\mathrm{re} \boldsymbol{\Theta}_\mathrm{t}\mathbf{H}_\mathrm{BR}\mathbf{w}_\mathrm{b}\right\|^2+\eta_0$},
	\item\emph{$	\widehat{\eta}_1=\frac{M\rho_\mathrm{re}}{P}\left\|\boldsymbol{\Omega}_\mathrm{re} \boldsymbol{\Theta}_\mathrm{t}\mathbf{H}_\mathrm{BR}\mathbf{w}_\mathrm{b}\right\|^2+\widehat{\eta}_0$}.
\end{itemize}

The existence of the Gamma functions in average secure rate $\widehat{R}^k_\mathrm{s}$ makes it challenging to be handled for solving the optimization problem in the next section. To tackle this issue, the lower bound of $\widehat{R}^k_\mathrm{s}$ is leveraged to replace $\widehat{R}^k_\mathrm{s}$ as a robust secure rate, which is expressed as
\begin{align}
    &\widehat{R}^k_\mathrm{s}>\widehat{R}^k_\mathrm{sl}=\notag\\&\begin{cases}
    	\left[\log_2\left(1+\gamma_{k0}\right)-\mathbb{E}_{\mathbf{h}_\mathrm{re}}\left(\log_2\left(1+\tilde{\gamma}_{\mathrm{e0}}\right)\right)\right]^+, &\mathcal{H}_0,\\
    	\left[\log_2\left(1+\gamma_{k1}\right)-\mathbb{E}_{\mathbf{h}_\mathrm{re}}\left(\log_2\left(1+\tilde{\gamma}_{\mathrm{e1}}\right)\right)\right]^+, &\mathcal{H}_1,
    \end{cases}	
\end{align}
where $\tilde{\gamma}_{\mathrm{e1}}=\frac{\left|\mathbf{h}^H_{\mathrm{re}}\boldsymbol{\Theta}_\mathrm{t}\mathbf{H}_\mathrm{BR}\mathbf{w}_k\right|^2}
{\left|\mathbf{h}^H_{\mathrm{re}}\boldsymbol{\Theta}_\mathrm{t}\mathbf{H}_\mathrm{BR}\mathbf{w}_b\right|^2+\sum\limits_{j\neq k}^K\left|\mathbf{h}^H_{\mathrm{re}}\boldsymbol{\Theta}_\mathrm{t}\mathbf{H}_\mathrm{BR}\mathbf{w}_j\right|^2}>\gamma_{\mathrm{e1}}$ and
$\tilde{\gamma}_{\mathrm{e0}}=\frac{\left|\mathbf{h}^H_{\mathrm{re}}\boldsymbol{\Theta}_\mathrm{t}\mathbf{H}_\mathrm{BR}\mathbf{w}_k\right|^2}
{\sum\limits_{j\neq k}^K\left|\mathbf{h}^H_{\mathrm{re}}\boldsymbol{\Theta}_\mathrm{t}\mathbf{H}_\mathrm{BR}\mathbf{w}_j\right|^2}>\gamma_{\mathrm{e0}}$.
%Based on the above analysis,
We can further derive that $\mathbb{E}_{\mathbf{h}_\mathrm{re}}\left(\log_2\left(1+\tilde{\gamma}_{\mathrm{e0}}\right)\right)=\log_2\left(\frac{\eta_0}{\widehat{\eta}_0}\right)$ $\mathbb{E}_{\mathbf{h}_\mathrm{re}}\left(\log_2\left(1+\tilde{\gamma}_{\mathrm{e1}}\right)\right)=\log_2\left(\frac{\eta_1}{\widehat{\eta}_1}\right)$.
Hence, the lower bound robust counterpart of the average secure rate for the $k$-the security user under two hypotheses are given by
\begin{align}
  \widehat{R}^k_\mathrm{sl, 0}=\left[\log_2\left(1+\gamma_{k0}\right)-\log_2\left(\frac{\eta_0}{\widehat{\eta}_0}\right)\right]^+,
\end{align}
\begin{align}
   \widehat{R}^k_\mathrm{sl, 1}=\left[\log_2\left(1+\gamma_{k1}\right)-\log_2\left(\frac{\eta_1}{\widehat{\eta}_1}\right)\right]^+,
\end{align}

which will be used in the next section for variables optimization and algorithm design.

%\vspace{-2mm}
\section{Problem Formulation and Algorithm Design}\label{sec:S4}
\subsection{Optimization Problem Formulation}
In this section, we will establish the optimization problem based on the theoretical analysis in Section \ref{sec:S3}.
Due to the fact that the existence of the CCs between BS and Bob is under a binary hypothesis, %not continuous in a time slot,
we define a Bernoulli variable $b$ %to represent the existence of the covert information in each slot time
where $b=0$ with the probability of $P_0$ means that BS only transmits the secure information, while $b=1$ with the probability of $P_1=1-P_0$ represents that BS transmits both the covert and secure messages.
In this paper, we maximize the average sum rate between the covert rate and the minimum secure rate over $b$ in a time slot while ensuring the covert constraint and the QoS constraints at Bob and $K$ security users by jointly optimizing the active and passive beamforming variables, i.e., $\mathbf{w}_\mathrm{b}$, $\{\mathbf{w}_k\}_{k=1}^K$, and $\boldsymbol{\Theta}_\mathrm{r}$, $\boldsymbol{\Theta}_\mathrm{t}$.
Specifically, the optimization objective of the average sum rate over the Bernoulli variable $b$ can be expressed as
\begin{align}\label{ASRate}
	%\overline{R}
&\overline{R}\big(\mathbf{w}_\mathrm{b}, \left\{\mathbf{w}_k\right\}_{k=1}^K, \boldsymbol{\Theta}_\mathrm{r}, \boldsymbol{\Theta}_\mathrm{t}\big) \notag\\
=&\mathbb{E}_{b}\Big(b R^\mathrm{c}_\mathrm{b}+L(b)\min_{k}\widehat{R}^k_\mathrm{sl, 0}+b\min_{k}\widehat{R}^k_\mathrm{sl, 1}\Big)\notag\\
	=&P_ 1R^\mathrm{c}_\mathrm{b}+P_ 0\min_{k}\widehat{R}^k_\mathrm{sl, 0}+P_ 1\min_{k}\widehat{R}^k_\mathrm{sl, 1},
\end{align}
where $L(\cdot)$ is the logical operator with $L(0)=1$, $L(1)=0$.

Based on the above analysis, the optimization problem is formulated as
\begin{subequations}\label{eq_orig_opti}
	\begin{align}
		&\max _{\mathbf{w}_\mathrm{b}, \left\{\mathbf{w}_k\right\}_{k=1}^K, \boldsymbol{\Theta}_\mathrm{r}, \boldsymbol{\Theta}_\mathrm{t} }
\overline{R}\big(\mathbf{w}_\mathrm{b}, \left\{\mathbf{w}_k\right\}_{k=1}^K, \boldsymbol{\Theta}_\mathrm{r}, \boldsymbol{\Theta}_\mathrm{t}\big),\notag \\
		&\quad~\text { s.t. }\left\|\mathbf{w}_\mathrm{b}\right\|_2^2+\sum\limits_{k=1}^K \left\|\mathbf{w}_k\right\|_2^2\leq P_\mathrm{tmax},\label{eq_orig_opti_1}\\
		&\qquad~~~~e^{-\frac{\beta\ln{\frac{\alpha+\beta}{\beta}}}{\alpha}}\left(1-\frac{\beta}{\alpha+\beta}\right)\leq\epsilon,\label{eq_orig_opti_2}\\
		&\qquad~~~~R^\mathrm{c}_\mathrm{b}\geq R_\mathrm{b}^*,\label{eq_orig_opti_3}\\
		&\qquad~~~~\min_{k}\widehat{R}^k_\mathrm{sl, 0}\geq R_\mathrm{s0}^*,~k\in\mathcal{K}, \label{eq_orig_opti_4}\\
		&\qquad~~~~\min_{k}\widehat{R}^k_\mathrm{sl, 1}\geq R_\mathrm{s1}^*,~k\in\mathcal{K}, \label{eq_orig_opti_5}\\
		&\qquad~~~~\beta_\mathrm{r}^m+\beta_\mathrm{t}^m=1,\phi_\mathrm{r}^m,~\phi_\mathrm{t}^m\in[0, 2\pi),~m\in\mathcal{M}, \label{eq_orig_opti_6}
	\end{align}
\end{subequations}
where \eqref{eq_orig_opti_1} is the transmit power constraint of the BS with $P_\mathrm{tmax}$ being the maximum power budget; \eqref{eq_orig_opti_2} denotes the covertness constraint, which is equivalent to $P_\mathrm{ea}^*\geq1-\epsilon$; \eqref{eq_orig_opti_3} and \eqref{eq_orig_opti_4}, \eqref{eq_orig_opti_5} represent the QoS constraints for covert rate and secure rate with the minimum required covert rate $R_\mathrm{b}^*$ and secure rate $R_\mathrm{s0}^*$ and $R_\mathrm{s1}^*$;  \eqref{eq_orig_opti_6} is the amplitude and phase shift constraints for STAR-RIS.
In fact, solving this optimization problem is quite challenging due to the strong coupling among variables, i.e., $\mathbf{w}_\mathrm{b}$, $\{\mathbf{w}_k\}_{k=1}^K$, $\boldsymbol{\Theta}_\mathrm{r}$ and $\boldsymbol{\Theta}_\mathrm{t}$, in the objective function, covert constraint and QoS constraints.
Additionally, the characteristic amplitude constraint introduced by STAR-RIS complicates the problem because $\boldsymbol{\Theta}_\mathrm{r}$ and $\boldsymbol{\Theta}_\mathrm{t}$ depend on each other in terms of element amplitudes. As a result, it is difficult to directly solve the optimization problem \eqref{eq_orig_opti} using convex optimization algorithms. To address this challenge, we propose an iterative algorithm that leverages an alternative strategy to effectively solve this optimization problem, which is presented in the next section.
\vspace{-2mm}
\subsection{Algorithm Design}
In this section, we detail the proposed iterative algorithm for solving the original formulated problem \eqref{eq_orig_opti}. Specifically, this problem is divided into two subproblems which are solved to the design active and passive beamformers, respectively.
\subsubsection{Joint Active beamforming design for $\mathbf{w}_\mathrm{b}$ and $\{\mathbf{w}_k\}_{k=1}^K$}
We first design the active beamforming variables $\mathbf{w}_\mathrm{b}$ and $\{\mathbf{w}_k\}_{k=1}^K$ with  given the passive beamforming variables, i.e., $\boldsymbol{\Theta}_\mathrm{r}$ and $\boldsymbol{\Theta}_\mathrm{t}$.
In this circumstance, the original optimization problem can be simplified as
\begin{subequations}\label{eq_active1}
	\begin{align}
		&\max _{\mathbf{w}_\mathrm{b}, \left\{\mathbf{w}_k\right\}_{k=1}^K} \overline{R}\big(\mathbf{w}_\mathrm{b}, \left\{\mathbf{w}_k\right\}_{k=1}^K\big),\notag \\
		&\quad~\text { s.t. }~~~\eqref{eq_orig_opti_1}-\eqref{eq_orig_opti_5}.\label{eq_active1_1}
	\end{align}
\end{subequations}

Problem \eqref{eq_active1} is still a non-convex optimization problem due to the non-convexity of the objective function, the covert constraint and the QoS constraints w.r.t. the active beamforming variables $\mathbf{w}_\mathrm{b}$ and $\left\{\mathbf{w}_k\right\}_{k=1}^K$. To tackle this problem, we first introduce three auxiliary variable $\iota$, $\kappa$ and $\varpi$ to replace $R^\mathrm{c}_\mathrm{b}$, $\min\limits_{k}\widehat{R}^k_\mathrm{sl, 1}$ and $\min\limits_{k}\widehat{R}^k_\mathrm{sl, 0}$ in the objective function and the QoS constraints \eqref{eq_orig_opti_3}-\eqref{eq_orig_opti_5}, which can transform the non-convex objective function into convex. In addition, it is easy to verify that the left-side of \eqref{eq_orig_opti_2} is a monotonically decreasing function of $\frac{\beta}{\alpha}$, and thus the covert constraint \eqref{eq_orig_opti_2} can be equivalently transformed as
\begin{align}\label{CC_simp}
  \frac{\beta}{\alpha}\geq \varphi(\epsilon),
\end{align}
where $\varphi(\epsilon)$ can be obtained by using the numerical methods such bisection search method.

Hence, problem \eqref{eq_active1} can be equivalently transformed into problem  \eqref{eq_active2} shown at the top of the next page,  where $\mathbf{D}=\sum\limits_{l=1}^{L}\boldsymbol{\Psi}^l_\mathrm{BR}\boldsymbol{\vartheta}_\mathrm{r}^T\boldsymbol{\Xi}^T\Big(\big(\widehat{\boldsymbol{\Psi}}^l_\mathrm{BR}\big)^T\otimes
\big(\boldsymbol{\Omega}_\mathrm{rw}^H\boldsymbol{\Omega}_\mathrm{rw}\big)\Big)\boldsymbol{\Xi}\boldsymbol{\vartheta}_\mathrm{r}^*$.
The convert constraint in \eqref{CC_simp} is re-expressed as \eqref{eq_active2_2} based on the definitions of $\alpha$ and $\beta$.
In \mbox{fact}, \eqref{eq_active2} is still a non-convex optimization problem because of the non-convexity of the constraints  \eqref{eq_active2_2}, \eqref{eq_active2_4}, \eqref{eq_active2_5} and \eqref{eq_active2_6}. To effectively address this problem, we resort to the SDR method \cite{luo2010semidefinite}. Specifically, we first let $\mathbf{W}=\left\{\mathbf{w}_\mathrm{b}, \mathbf{w}_1, \mathbf{w}_2, \cdots, \mathbf{w}_K\right\}$
and $\mathbf{W}_\mathrm{cs}=\operatorname{vec}\left(\mathbf{W}\right)\operatorname{vec}\left(\mathbf{W}\right)^H$, then the optimization problem \eqref{eq_active2} can be equivalently transformed as \eqref{eq_active3}, which is presented in the next page and where
\begin{figure*}[t]
		\hrulefill
\begin{subequations}\label{eq_active2}
		\begin{align}
			&\max _{\mathbf{w}_\mathrm{b}, \left\{\mathbf{w}_k\right\}_{k=1}^K, \iota, \kappa, \varpi} P_1\iota+P_1\kappa+P_0\varpi,\notag \\
			&\qquad~~\text { s.t. }\left\|\mathbf{w}_\mathrm{b}\right\|_2^2+\sum\limits_{k=1}^K \left\|\mathbf{w}_k\right\|_2^2\leq P_\mathrm{tmax},\label{eq_active2_1}\\			&\qquad\qquad~~\sum\limits_{k=1}^K\left(\mathbf{w}_k^H\mathbf{D}\mathbf{w}_k\right)\geq\left(\mathbf{w}_\mathrm{b}^H\mathbf{D}\mathbf{w}_\mathrm{b}\right)\varphi(\epsilon),\label{eq_active2_2}\\
			&\qquad\qquad~~\iota\geq R_\mathrm{b}^*, \varpi\geq R_\mathrm{s0}^*, \kappa\geq R_\mathrm{s1}^*,\label{eq_active2_3}\\
				&\qquad\qquad~~\log_2\left(1+\frac{\left|\mathbf{h}^H_{\mathrm{rb}}\boldsymbol{\Theta}_\mathrm{r}\mathbf{H}_\mathrm{BR}\mathbf{w}_\mathrm{b}\right|^2}{\sum_{k=1}^K\left|\mathbf{h}^H_{\mathrm{rb}}\boldsymbol{\Theta}_\mathrm{r}\mathbf{H}_\mathrm{BR}\mathbf{w}_k\right|^2+\sigma_\mathrm{b}^2}\right)\geq \iota,\label{eq_active2_4}\\
			&\qquad\qquad~~\log_2\left(1+\frac{\left|\mathbf{h}^H_{\mathrm{r}k}\boldsymbol{\Theta}_\mathrm{t}\mathbf{H}_\mathrm{BR}\mathbf{w}_k\right|^2}
			{\left\|\mathbf{h}^H_{\mathrm{r}k}\boldsymbol{\Theta}_\mathrm{t}\mathbf{H}_\mathrm{BR}\mathbf{W}_{-k}\right\|^2_2+\sigma_{\mathrm{r} k}^2}\right)-
			 \log_2\left(1+\frac{\left\|\boldsymbol{\Omega}_\mathrm{re}\boldsymbol{\Theta}_\mathrm{t}\mathbf{H}_\mathrm{BR}\mathbf{w}_k\|\right._2^2}{\left\|\boldsymbol{\Omega}_\mathrm{re}\boldsymbol{\Theta}_\mathrm{t}\mathbf{H}_\mathrm{BR}\mathbf{W}_{-k}\|\right.^2_\mathrm{F}}\right)\geq \varpi ,~ \forall k,\label{eq_active2_5}\\
				&\qquad\qquad~~ \log_2\left(1+\frac{\left|\mathbf{h}^H_{\mathrm{r}k}\boldsymbol{\Theta}_\mathrm{t}\mathbf{H}_\mathrm{BR}\mathbf{w}_k\right|^2}
			{\left|\mathbf{h}^H_{\mathrm{r}k}\boldsymbol{\Theta}_\mathrm{t}\mathbf{H}_\mathrm{BR}\mathbf{w}_\mathrm{b}\right|^2+\left\|\mathbf{h}^H_{\mathrm{r}k}\boldsymbol{\Theta}_\mathrm{t}\mathbf{H}_\mathrm{BR}\mathbf{W}_{-k}\right\|^2_2+\sigma_{k}^2}\right)-\notag\\
	&\qquad\qquad~~\log_2\left(1+\frac{\left\|\boldsymbol{\Omega}_\mathrm{re}\boldsymbol{\Theta}_\mathrm{t}\mathbf{H}_\mathrm{BR}\mathbf{w}_k\|\right._2^2}{\left\|\boldsymbol{\Omega}_\mathrm{re}\boldsymbol{\Theta}_\mathrm{t}\mathbf{H}_\mathrm{BR}\mathbf{w}_\mathrm{b}\|\right._2^2+		\left\|\boldsymbol{\Omega}_\mathrm{re}\boldsymbol{\Theta}_\mathrm{t}\mathbf{H}_\mathrm{BR}\mathbf{W}_{-k}\|\right.^2_\mathrm{F}}\right)
			\geq \kappa,~\forall k. \label{eq_active2_6}
		\end{align}
	\end{subequations}
\end{figure*}
 \begin{figure*}[t]
 	\hrulefill
\begin{subequations}\label{eq_active3}
	\begin{align}
		&\max _{\mathbf{W}_\mathrm{cs}, \iota, \kappa, \varpi} P_1\iota+P_1\kappa+P_0\varpi,\notag \\
		&~~\text { s.t. }\operatorname{Tr}(\mathbf{W}_\mathrm{cs})\leq P_\mathrm{tmax},\label{eq_active3_1}\\
		&\qquad~~\operatorname{Tr}(\mathbf{W}_\mathrm{cs}\widehat{\mathbf{D}})\geq\operatorname{Tr}(\mathbf{W}_\mathrm{cs}\mathbf{D}_1)\varphi(\epsilon),\label{eq_active3_2}\\
		&\qquad~~\eqref{eq_active2_2},\label{eq_active3_3}\\
		&\qquad~~f(\mathbf{W}_\mathrm{cs})=\log_2\left(\operatorname{Tr}(\mathbf{W}_\mathrm{cs}\widehat{\mathbf{A}})+\sigma_\mathrm{b}^2\right)-\log_2\left(\operatorname{Tr}(\mathbf{W}_\mathrm{cs}\tilde{\mathbf{A}})+\sigma_\mathrm{b}^2\right)\geq\iota,\label{eq_active3_4}\\
			&\qquad~~ f_{k, 1}(\mathbf{W}_\mathrm{cs})= \nonumber\\
			&\qquad~~\log_2\left(\operatorname{Tr}(\mathbf{W}_\mathrm{cs}\check{\mathbf{B}}_k)+\sigma_{k}^2\right)+\log_2\left(\operatorname{Tr}(\mathbf{W}_\mathrm{cs}\breve{\mathbf{C}}_k)\right)-\log_2\left(\operatorname{Tr}(\mathbf{W}_\mathrm{cs}\breve{\mathbf{B}}_k)+\sigma_{k}^2\right)-\log_2\left(\operatorname{Tr}(\mathbf{W}_\mathrm{cs}\check{\mathbf{C}})\right)\geq \varpi ,~ \forall k,\label{eq_active3_5}\\
			&\qquad~~f_{k, 2}(\mathbf{W}_\mathrm{cs})= \nonumber\\
			&\qquad~~\log_2\left(\operatorname{Tr}(\mathbf{W}_\mathrm{cs}\widehat{\mathbf{B}}_k)+\sigma_{k}^2\right)+\log_2\left(\operatorname{Tr}(\mathbf{W}_\mathrm{cs}\tilde{\mathbf{C}}_k)\right)-\log_2\left(\operatorname{Tr}(\mathbf{W}_\mathrm{cs}\tilde{\mathbf{B}}_k)+\sigma_{k}^2\right)-\log_2\left(\operatorname{Tr}(\mathbf{W}_\mathrm{cs}\widehat{\mathbf{C}})\right)\geq \kappa , ~\forall k,\label{eq_active3_6}\\
		&\qquad~~ \mathbf{W}_\mathrm{cs}\succeq 0,~\operatorname{rank}(\mathbf{W}_\mathrm{cs})=1.\label{eq_active3_7}
	\end{align}
\end{subequations}
 	\hrulefill
\end{figure*}
\begin{itemize}
	\item \emph{$\widehat{\mathbf{D}}=\big(\mathbf{E}_{-1}\mathbf{E}_{-1}^T\big)\otimes \mathbf{D},~\mathbf{D}_1=\big(\mathbf{e}_{1}\mathbf{e}_{1}^T\big)\otimes \mathbf{D}$},
\item \emph{$\widehat{\mathbf{A}}=\mathbf{I}_{K+1}\otimes \mathbf{A},~\tilde{\mathbf{A}}=\big(\mathbf{E}_{-1}\mathbf{E}_{-1}^T\big)\otimes \mathbf{A}$},
\item \emph{$\check{\mathbf{B}}_k=\big(\mathbf{E}_{-1}\mathbf{E}_{-1}^T\big)\otimes \mathbf{B}_k,~\tilde{\mathbf{B}}_k=\big(\mathbf{E}_{-(k+1)}\mathbf{E}_{-(k+1)}^T\big)\otimes \mathbf{B}_k$},
\item \emph{$\breve{\mathbf{B}}_k=\big(\mathbf{E}_{-(1, k+1)}\mathbf{E}_{-(1, k+1)}^T\big)\otimes \mathbf{B}_k,~\widehat{\mathbf{B}}_k=\mathbf{I}_{K+1}\otimes \mathbf{B}_k$},
\item \emph{$\check{\mathbf{C}}_k=\big(\mathbf{E}_{-1}\mathbf{E}_{-1}^T\big)\otimes \mathbf{C},~\tilde{\mathbf{C}}_k=\big(\mathbf{E}_{-(k+1)}\mathbf{E}_{-(k+1)}^T\big)\otimes \mathbf{C}$},
\item \emph{$\breve{\mathbf{C}}_k=\big(\mathbf{E}_{-(1, k+1)}\mathbf{E}_{-(1, k+1)}^T\big)\otimes \mathbf{C},~\widehat{\mathbf{C}}=\mathbf{I}_{K+1}\otimes \mathbf{C}$},
\item \emph{$\mathbf{E}_{-(1, k+1)}=\big\{\mathbf{e}_2, \cdots, \mathbf{e}_{k},\mathbf{e}_{k+2}, \cdots, \mathbf{e}_{K+1}\big\}$},
\item \emph{$\mathbf{E}_{-1}=\big\{\mathbf{e}_2, \cdots, \mathbf{e}_{k}, \cdots, \mathbf{e}_{K+1}\big\}$},
\item \emph{$\mathbf{E}_{-(k+1)}=\big\{\mathbf{e}_1, \cdots, \mathbf{e}_{k}, \mathbf{e}_{k+2}, \cdots, \mathbf{e}_{K+1}\big\}$},
\item \emph{$\mathbf{A}=\left(\mathbf{h}_\mathrm{rb}^H\boldsymbol{\Theta}_\mathrm{r}\mathbf{H}_\mathrm{BR}\right)^H\left(\mathbf{h}_\mathrm{rb}^H\boldsymbol{\Theta}_\mathrm{r}\mathbf{H}_\mathrm{BR}\right)$},
\item \emph{$\mathbf{B}_k=\left(\mathbf{h}_{\mathrm{r}k}^H\boldsymbol{\Theta}_\mathrm{t}\mathbf{H}_\mathrm{BR}\right)^H\left(\mathbf{h}_{\mathrm{r}k}^H\boldsymbol{\Theta}_\mathrm{t}\mathbf{H}_\mathrm{BR}\right)$},
\item \emph{$	\mathbf{C}=\mathbf{H}_\mathrm{BR}^H\boldsymbol{\Theta}_\mathrm{t}^H\boldsymbol{\Omega}_\mathrm{re}^H\boldsymbol{\Omega}_\mathrm{re}\boldsymbol{\Theta}_\mathrm{t}\mathbf{H}_\mathrm{BR}$}.
\end{itemize}

Note that problem \eqref{eq_active3} is still a non-convex optimization problem due to the non-convex constraints \eqref{eq_active3_4}, \eqref{eq_active3_5}, \eqref{eq_active3_6} and the rank-one constraint in \eqref{eq_active3_7}.
To transform \eqref{eq_active3} into a solvable convex problem, we first handle the constraints \eqref{eq_active3_4}, \eqref{eq_active3_5} and \eqref{eq_active3_6}.
In particular, we can find that $f(\mathbf{W}_\mathrm{cs})$, $f_{k, 1}(\mathbf{W}_\mathrm{cs})$ and $f_{k, 2}(\mathbf{W}_\mathrm{cs})$ are all difference of concave (DC) functions, and thus the first-order Taylor expansion can be  leveraged on them to obtain their concave lower bounds  in the $i$-th innerloop iteration of the proposed iterative algorithm (See Algorithm 1 in Section \ref{Algorithm} ). These concave lower bounds will be adopted to replace the original expressions in the optimization problem \eqref{eq_active3} which are derived as
\vspace{-2mm}\begin{align} f(\mathbf{W}_\mathrm{cs})&\geq\log_2\Big(\operatorname{Tr}\big(\mathbf{W}_\mathrm{cs}\widehat{\mathbf{A}}\big)+\sigma_\mathrm{b}^2\Big)-g_1\big(\mathbf{W}_\mathrm{cs}, \mathbf{W}_\mathrm{cs}^{(i)}\big)\notag\\
	&\triangleq\widehat{f}\big(\mathbf{W}_\mathrm{cs}, \mathbf{W}_\mathrm{cs}^{(i)}\big),\notag\\
f_{k, 1}(\mathbf{W}_\mathrm{cs})&\geq\log_2\Big(\operatorname{Tr}(\mathbf{W}_\mathrm{cs}\check{\mathbf{B}}_k)+\sigma_{k}^2\Big)-g_{k, 1}\big(\mathbf{W}_\mathrm{cs}, \mathbf{W}_\mathrm{cs}^{(i)}\big)\notag\\
&+\log_2\Big(\operatorname{Tr}\big(\mathbf{W}_\mathrm{cs}\breve{\mathbf{C}}_k\big)\Big)-g_2\big(\mathbf{W}_\mathrm{cs}, \mathbf{W}_\mathrm{cs}^{(i)}\big)\notag\\
&\triangleq\widehat{f}_{k, 1}\big(\mathbf{W}_\mathrm{cs}, \mathbf{W}_\mathrm{cs}^{(i)}\big),\\
	f_{k, 2}(\mathbf{W}_\mathrm{cs})&\geq \log_2\Big(\operatorname{Tr}\big(\mathbf{W}_\mathrm{cs}\widehat{\mathbf{B}}_k\big)+\sigma_{k}^2\Big)-g_{k, 2}\big(\mathbf{W}_\mathrm{cs}, \mathbf{W}_\mathrm{cs}^{(i)}\big)\notag\\
	&+\log_2\Big(\operatorname{Tr}\big(\mathbf{W}_\mathrm{cs}\tilde{\mathbf{C}}_k\big)\Big)
	-g_3\big(\mathbf{W}_\mathrm{cs}, \mathbf{W}_\mathrm{cs}^{(i)}\big)\notag\\
	&\triangleq\widehat{f}_{k, 2}\big(\mathbf{W}_\mathrm{cs}, \mathbf{W}_\mathrm{cs}^{(i)}\big),
\end{align}
where the expressions of $g_1\big(\mathbf{W}_\mathrm{cs}, \mathbf{W}_\mathrm{cs}^{(i)}\big)$, $g_2\big(\mathbf{W}_\mathrm{cs}, \mathbf{W}_\mathrm{cs}^{(i)}\big)$, $g_3\big(\mathbf{W}_\mathrm{cs}, \mathbf{W}_\mathrm{cs}^{(i)}\big)$, and $g_{k, 1}\big(\mathbf{W}_\mathrm{cs}, \mathbf{W}_\mathrm{cs}^{(i)}\big)$, $g_{k, 2}\big(\mathbf{W}_\mathrm{cs}, \mathbf{W}_\mathrm{cs}^{(i)}\big)$ for $k\in\mathcal{K}$ are given in \eqref{eq_taylor_expansion}.
\begin{subequations}\label{eq_taylor_expansion}
	\begin{figure*}[t]
		\hrulefill
		\begin{align}
			g_1\big(\mathbf{W}_\mathrm{cs}, \mathbf{W}_\mathrm{cs}^{(t)}\big)=&\log_2\Big(\operatorname{Tr}(\mathbf{W}_\mathrm{cs}^{(i)}\tilde{\mathbf{A}})+\sigma_\mathrm{b}^2\Big)+
			\frac{\operatorname{Tr}(\mathbf{W}_\mathrm{cs}\tilde{\mathbf{A}})-\operatorname{Tr}(\mathbf{W}_\mathrm{cs}^{(i)}\tilde{\mathbf{A}})}{\ln2\Big(\operatorname{Tr}(\mathbf{W}_\mathrm{cs}^{(i)}\tilde{\mathbf{A}})+\sigma_\mathrm{b}^2\Big)},\\
			g_{k, 1}\big(\mathbf{W}_\mathrm{cs}, \mathbf{W}_\mathrm{cs}^{(i)}\big)=&\log_2\left(\operatorname{Tr}(\mathbf{W}_\mathrm{cs}\breve{\mathbf{B}}_k)+\sigma_{k}^2\right)+\frac{\operatorname{Tr}(\mathbf{W}_\mathrm{cs}\breve{\mathbf{B}}_k)-\operatorname{Tr}(\mathbf{W}_\mathrm{cs}^{(i)}\breve{\mathbf{B}}_k)}{\ln2\left(\operatorname{Tr}(\mathbf{W}_\mathrm{cs}^{(i)}\breve{\mathbf{B}}_k)+\sigma_{k}^2\right)},\\
			g_2\big(\mathbf{W}_\mathrm{cs}, \mathbf{W}_\mathrm{cs}^{(i)}\big)=&\frac{\operatorname{Tr}(\mathbf{W}_\mathrm{cs}\check{\mathbf{C}})-\operatorname{Tr}(\mathbf{W}_\mathrm{cs}^{(i)}\check{\mathbf{C}})}{\ln2\left(\operatorname{Tr}(\mathbf{W}_\mathrm{cs}^{(i)}\check{\mathbf{C}})\right)}+\log_2\left(\operatorname{Tr}(\mathbf{W}_\mathrm{cs}^{(i)}\check{\mathbf{C}})\right),\\
			g_{k, 2}\big(\mathbf{W}_\mathrm{cs}, \mathbf{W}_\mathrm{cs}^{(i)}\big)=&\log_2\left(\operatorname{Tr}(\mathbf{W}_\mathrm{cs}\tilde{\mathbf{B}}_k)+\sigma_{k}^2\right)+\frac{\operatorname{Tr}(\mathbf{W}_\mathrm{cs}\tilde{\mathbf{B}}_k)-\operatorname{Tr}(\mathbf{W}_\mathrm{cs}^{(i)}\tilde{\mathbf{B}}_k)}{\ln2\left(\operatorname{Tr}(\mathbf{W}_\mathrm{cs}^{(i)}\tilde{\mathbf{B}}_k)+\sigma_{k}^2\right)},\\
			g_3\big(\mathbf{W}_\mathrm{cs}, \mathbf{W}_\mathrm{cs}^{(i)}\big)=&\frac{\operatorname{Tr}(\mathbf{W}_\mathrm{cs}\widehat{\mathbf{C}})-\operatorname{Tr}(\mathbf{W}_\mathrm{cs}^{(i)}\widehat{\mathbf{C}})}{\ln2\left(\operatorname{Tr}(\mathbf{W}_\mathrm{cs}^{(i)}\widehat{\mathbf{C}})\right)}+\log_2\left(\operatorname{Tr}(\mathbf{W}_\mathrm{cs}^{(i)}\widehat{\mathbf{C}})\right).
		\end{align}
	\end{figure*}
\end{subequations}

For the rank-one constraint in \eqref{eq_active3_7}, we choose to equivalently rewrite it as \cite{xiao2023simultaneously}
\vspace{-2mm}\begin{align}
\operatorname{rank}(\mathbf{W}_\mathrm{cs})=1 \Leftrightarrow \operatorname{Tr}(\mathbf{W}_\mathrm{cs}) -\left\|\mathbf{W}_\mathrm{cs}\right\|_2=0,
\end{align}
where $\left\|\mathbf{W}_\mathrm{cs}\right\|_2$ denotes the spectral norm  and is a convex function w.r.t. $\mathbf{W}_\mathrm{cs}$. It is worth noting that for any positive semidefinite matrix $\mathbf{A}$, $\eta_\mathrm{cs}(\mathbf{A})\triangleq\operatorname{Tr}(\mathbf{A}) -\left\|\mathbf{A}\right\|_2\geq 0$ always holds and the equality is satisfied if and only if $\operatorname{rank}(\mathbf{A})=1$. Thus, based on the nonnegative characteristic of $\eta_\mathrm{cs}(\mathbf{W}_\mathrm{cs})$, we add it into the objective function as a penalty term  for the rank-one constraint which is subtracted by the objective function.
However, the objective function with the penalty term is non-concave and cannot be addressed by convex optimization algorithms directly. To tackle this issue, the spectral norm in $\eta_\mathrm{cs}(\mathbf{W}_\mathrm{cs})$ is replaced by its linear lower bound obtained by its first-order Taylor expansion. Hence, we can obtain the upper bound of $\eta_\mathrm{cs}(\mathbf{W}_\mathrm{cs})$, which is expressed as
\begin{align}\label{eq_rank-one_rewrite}
	&\eta_\mathrm{cs}(\mathbf{W}_\mathrm{cs})\leq \widehat{\eta}_\mathrm{cs}(\mathbf{W}_\mathrm{cs})\triangleq\operatorname{Tr}(\mathbf{W}_\mathrm{cs})-\notag\\
	&\Big(\|\mathbf{W}_\mathrm{cs}^{(i)}\|_2
	+\operatorname{Tr}\big(\mathbf{w}_\mathrm{cs}^{(i)}\big(\mathbf{w}_\mathrm{cs}^{(i)}\big)^H\big(\mathbf{W}_\mathrm{cs}-\mathbf{W}_\mathrm{cs}^{(i)}\big)\big)\Big),
\end{align}
where $\mathbf{w}_\mathrm{cs}^{(i)}$ represents the eigenvectors corresponding to the largest eigenvalues of $\mathbf{W}_\mathrm{cs}^{(i)}$ in $i$-th inner loop iteration. Thus, the objective function with $\widehat{\eta}_\mathrm{cs}(\mathbf{W}_\mathrm{cs})$ will be adopted to calculate the $(i+1)$-th solution, denoted as $\mathbf{W}_\mathrm{cs}^{(i+1)}$. According to the above analysis, the optimization problem \eqref{eq_active3} can be further transformed as
  \begin{subequations}\label{eq_active4}
  	\begin{align}
  		&\max _{\mathbf{W}_\mathrm{cs}, \iota, \kappa, \varpi} P_1\iota+P_1\kappa+P_0\varpi-\varrho_\mathrm{cs}\widehat{\eta}_\mathrm{cs}(\mathbf{W}_\mathrm{cs}),\notag \\
  		&~~\text { s.t. }~\eqref{eq_active3_1}, \eqref{eq_active3_2}, \eqref{eq_active3_3}, \label{eq_active4_1}\\
  		&\qquad~~\widehat{f}\big(\mathbf{W}_\mathrm{cs}, \mathbf{W}_\mathrm{cs}^{(i)}\big)\geq\iota,\label{eq_active4_2}\\
   		&\qquad~~\widehat{f}_{k, 1}\big(\mathbf{W}_\mathrm{cs}, \mathbf{W}_\mathrm{cs}^{(i)}\big)\geq \varpi , ~\forall k,\label{eq_active4_3}\\
  		&\qquad~~\widehat{f}_{k, 2}\big(\mathbf{W}_\mathrm{cs}, \mathbf{W}_\mathrm{cs}^{(i)}\big)\geq \kappa , ~\forall k,\label{eq_active4_4}\\
  		&\qquad~~ \mathbf{W}_\mathrm{cs}\succeq 0,\label{eq_active4_5}
  	\end{align}
  \end{subequations}
where $\varrho_\mathrm{cs}$ is the penalty coefficient. The optimization problem \eqref{eq_active4} is a standard convex semidefinite programming (SDP) problem which is able to be effectively solved by the existing convex optimization tools such as CVX \cite{grant14cvx}.
\subsubsection{Joint Passive beamforming design for $\boldsymbol{\Theta}_\mathrm{r}$ and $\boldsymbol{\Theta}_\mathrm{t}$}
After obtaining the active beamformers, we then design the passive beamforming variables $\boldsymbol{\Theta}_\mathrm{r}$ and $\boldsymbol{\Theta}_\mathrm{t}$ with  given the obtained $\mathbf{w}_\mathrm{b}$ and $\left\{\mathbf{w}_k\right\}_{k=1}^K$.
Specifically, based on the original optimization problem \eqref{eq_orig_opti} and definition of the average sum rate in \eqref{ASRate}, the optimization problem for joint designing the passive beamforming variables $\boldsymbol{\Theta}_\mathrm{r}$ and $\boldsymbol{\Theta}_\mathrm{t}$ can be expressed as
\begin{subequations}\label{eq_passive1}
     \begin{align}
    	&\max _{\boldsymbol{\Theta}_\mathrm{r}, \boldsymbol{\Theta}_\mathrm{t}} ~~~\overline{R}\big(\boldsymbol{\Theta}_\mathrm{r}, \boldsymbol{\Theta}_\mathrm{t}\big),\notag \\
    	&~\text { s.t. } ~~~\eqref{eq_orig_opti_2}-\eqref{eq_orig_opti_6}.\label{eq_passive1_1}
     \end{align}
 \end{subequations}

\vspace{-2mm}Note that problem \eqref{eq_passive1} is a non-convex optimization problem w.r.t. $\boldsymbol{\Theta}_\mathrm{r}$ and $ \boldsymbol{\Theta}_\mathrm{t}$. Similarly, we will adopt the SDR techniques to deal with this optimization problem.
The reformulated covert constraint $\frac{\beta}{\alpha}\geq \varphi(\epsilon)$ in \eqref{eq_active2_2} is still utilized to guarantee the covert performance.
Let $\mathbf{Q}_\mathrm{r}=\boldsymbol{\vartheta}_\mathrm{r}^*\boldsymbol{\vartheta}_\mathrm{r}^T$, $\mathbf{Q}_\mathrm{t}=\boldsymbol{\vartheta}_\mathrm{t}^*\boldsymbol{\vartheta}_\mathrm{t}^T$ where $\boldsymbol{\vartheta}_\mathrm{r}=\operatorname{diag}(\boldsymbol{\Theta}_\mathrm{r})$, $\boldsymbol{\vartheta}_\mathrm{t}=\operatorname{diag}(\boldsymbol{\Theta}_\mathrm{t})$, and then
the optimization problem \eqref{eq_passive1} can be equivalently reformulated as problem \eqref{eq_passive2},
 \begin{subequations}\label{eq_passive2}
	\begin{align}
		&\hspace{-2mm}\max _{\mathbf{V}}~ P_1\iota+P_1\kappa+P_0\varpi,\notag \\
		&\hspace{-4mm}~~\text { s.t. } \operatorname{Tr}(\mathbf{Q}_\mathrm{r}\mathbf{F})\geq\operatorname{Tr}(\mathbf{Q}_\mathrm{r}\mathbf{E})\varphi(\epsilon),\label{eq_passive2_1}\\
		&\hspace{-4mm}\qquad~~\eqref{eq_active2_2},\label{eq_passive2_2}\\
		&\hspace{-4mm}\qquad~~\log_2\left(\operatorname{Tr}(\mathbf{Q}_\mathrm{r}\mathbf{G})+\operatorname{Tr}(\mathbf{Q}_\mathrm{r}\mathbf{O})+\sigma_\mathrm{b}\right)
		-\notag\\
		&\hspace{-4mm}\qquad~~\log_2\left(\operatorname{Tr}(\mathbf{Q}_\mathrm{r}\mathbf{O})+\sigma_\mathrm{b}\right)\geq \iota,\label{eq_passive2_3}\\
		&\hspace{-4mm}\qquad~~\log_2\left(\operatorname{Tr}(\mathbf{Q}_\mathrm{t}\mathbf{P}_k)+\sigma_{k}^2\right)-\log_2\left(\operatorname{Tr}(\mathbf{Q}_\mathrm{t}\check{\mathbf{P}}_k)+\sigma_{k}^2\right)\notag\\
		&\hspace{-4mm}\qquad~~-\log_2\left(\operatorname{Tr}(\mathbf{Q}_\mathrm{t}\mathbf{S})\right)+\log_2\left(\operatorname{Tr}\left(\mathbf{Q}_\mathrm{t}\check{\mathbf{S}}_k\right)\right)\geq \varpi,~\forall{k},\label{eq_passive2_4}\\
		&\hspace{-4mm}\qquad~~\log_2\left(\operatorname{Tr}(\mathbf{Q}_\mathrm{t}\mathbf{T}_k)+\sigma_{k}^2\right)-\log_2\left(\operatorname{Tr}(\mathbf{Q}_\mathrm{t}\check{\mathbf{T}}_k)+\sigma_{k}^2\right)\notag\\
		&\hspace{-4mm}\qquad~~-\log_2\left(\operatorname{Tr}(\mathbf{Q}_\mathrm{t}\mathbf{U})\right)+\log_2\left(\operatorname{Tr}(\mathbf{Q}_\mathrm{t}\check{\mathbf{U}}_k)\right)\geq\kappa,~\forall{k},\label{eq_passive2_5}\\
&\hspace{-4mm}\qquad~~\operatorname{diag}(\mathbf{Q}_\mathrm{r})=\boldsymbol{\beta}_\mathrm{r},\label{eq_passive2_6} \operatorname{diag}(\mathbf{Q}_\mathrm{t})=\boldsymbol{\beta}_\mathrm{t}\\
		&\hspace{-4mm}\qquad~~ \boldsymbol{\beta}_\mathrm{r}+\boldsymbol{\beta}_\mathrm{t}=\mathbf{I}_{M\times1},\label{eq_passive2_7}\\
		&\hspace{-4mm}\qquad~~\mathbf{Q}_{\mathrm{r}} \succeq 0, \mathbf{Q}_{\mathrm{t}} \succeq 0, \label{eq_passive2_8}\\
		&\hspace{-4mm}\qquad~~\operatorname{rank}(\mathbf{Q}_\mathrm{r})=1, \operatorname{rank}(\mathbf{Q}_\mathrm{t})=1, \label{eq_passive2_9}
	\end{align}
\end{subequations}
where
\begin{itemize}
	\item \emph{$\mathbf{V}=\{\mathbf{Q}_\mathrm{r}, \mathbf{Q}_\mathrm{t}, \boldsymbol{\beta}_\mathrm{r}, \boldsymbol{\beta}_\mathrm{t}, \iota, \kappa, \varpi\}$}is the defined optimization variable set,
	\item \emph{$\boldsymbol{\beta}_\mathrm{r}=\{\beta_\mathrm{r}^1, \cdots, \beta_\mathrm{r}^M \}$, $\boldsymbol{\beta}_\mathrm{t}=\{\beta_\mathrm{t}^1, \cdots, \beta_\mathrm{t}^M \}$},
	\item \emph{$\mathbf{E}=\sum_{l=1}^{L}\big(\mathbf{w}^H_\mathrm{b}\boldsymbol{\Psi}^l_\mathrm{BR}\mathbf{w}_\mathrm{b}\big)
	\boldsymbol{\Delta}^l$},
		\item \emph{$\mathbf{F}=\sum_{k=1}^K\sum_{l=1}^{L}\big(\mathbf{w}^H_k\boldsymbol{\Psi}^l_\mathrm{BR}\mathbf{w}_k\big)
	\boldsymbol{\Delta}^l$},
	\item \emph{$\boldsymbol{\Delta}^l=\boldsymbol{\Xi}^T\Big(\big(\widehat{\boldsymbol{\Psi}}^l_\mathrm{BR}\big)^T\otimes
	\big(\boldsymbol{\Omega}_\mathrm{rw}^H\boldsymbol{\Omega}_\mathrm{rw}\big)\Big)\boldsymbol{\Xi}$},
	 	\item \emph{$\mathbf{G}=\mathbf{H}_\mathrm{rb}^*\mathbf{H}_\mathrm{BR}\mathbf{w}_\mathrm{b}\mathbf{w}_\mathrm{b}^H\mathbf{H}_\mathrm{BR}^H\mathbf{H}_\mathrm{rb}^T$}, 	\item \emph{$\mathbf{O}=\mathbf{H}_\mathrm{rb}^*\mathbf{H}_\mathrm{BR}\sum_{k=1}^K\big(\mathbf{w}_k\mathbf{w}_k^H\big)\mathbf{H}_\mathrm{BR}^H\mathbf{H}_\mathrm{rb}^T$},
	\item \emph{$\mathbf{P}_k=\mathbf{H}_{\mathrm{r}k}^*\mathbf{H}_\mathrm{BR}\sum_{k=1}^K\big(\mathbf{w}_k\mathbf{w}_k^H\big)\mathbf{H}_\mathrm{BR}^H\mathbf{H}_{\mathrm{r}k}^T$},
	 	\item \emph{$\hat{\mathbf{P}}_k=\mathbf{H}_{\mathrm{r}k}^*\mathbf{H}_\mathrm{BR}\sum_{j\neq k}^K\big(\mathbf{w}_j\mathbf{w}_j^H\big)\mathbf{H}_\mathrm{BR}^H\mathbf{H}_{\mathrm{r}k}^T$},
		\item \emph{$\mathbf{S}=\sum_{k=1}^K\boldsymbol{\Xi}^T\Big(\big(\boldsymbol{\Omega}_\mathrm{re}^H\boldsymbol{\Omega}_\mathrm{re} \big)^T \otimes\big(\mathbf{H}_\mathrm{BR}\mathbf{w}_k\mathbf{w}_k^H\mathbf{H}_\mathrm{BR}^H\big)\Big)\boldsymbol{\Xi}$},
		\item \emph{$\hat{\mathbf{S}}_k=\sum_{j\neq k}^K\boldsymbol{\Xi}^T\Big(\big(\boldsymbol{\Omega}_\mathrm{re}^H\boldsymbol{\Omega}_\mathrm{re} \big)^T \otimes\big(\mathbf{H}_\mathrm{BR}\mathbf{w}_j\mathbf{w}_j^H\mathbf{H}_\mathrm{BR}^H\big)\Big)\boldsymbol{\Xi}$},
	\item \emph{$\mathbf{T}_k=\mathbf{H}_{\mathrm{r}k}^*\mathbf{H}_\mathrm{BR}\mathbf{w}_\mathrm{b}\mathbf{w}_\mathrm{b}^H
	\mathbf{H}_\mathrm{BR}^H\mathbf{H}_{\mathrm{r}k}^T+\mathbf{P}_k$},
		\item \emph{$\hat{\mathbf{T}}_k=\mathbf{H}_{\mathrm{r}k}^*\mathbf{H}_\mathrm{BR}\mathbf{w}_\mathrm{b}\mathbf{w}_\mathrm{b}^H
	\mathbf{H}_\mathrm{BR}^H\mathbf{H}_{\mathrm{r}k}^T+ \hat{\mathbf{P}}_k$},
		\item \emph{$\mathbf{U}=\boldsymbol{\Xi}^T\Big(\big(\boldsymbol{\Omega}_\mathrm{re}^H\boldsymbol{\Omega}_\mathrm{re} \big)^T \otimes\big(\mathbf{H}_\mathrm{BR}\mathbf{w}_\mathrm{b}\mathbf{w}_\mathrm{b}^H\mathbf{H}_\mathrm{BR}^H\big)\Big)\boldsymbol{\Xi}+\mathbf{S}$},
		\item \emph{$\hat{\mathbf{U}}_k=\boldsymbol{\Xi}^T\Big(\big(\boldsymbol{\Omega}_\mathrm{re}^H\boldsymbol{\Omega}_\mathrm{re} \big)^T \otimes\big(\mathbf{H}_\mathrm{BR}\mathbf{w}_\mathrm{b}\mathbf{w}_\mathrm{b}^H\mathbf{H}_\mathrm{BR}^H\big)\Big)\boldsymbol{\Xi}+	\hat{\mathbf{S}}_k$},
		\item \emph{$\mathbf{H}_\mathrm{rb}=\operatorname{Diag}(\mathbf{h}_\mathrm{rb}), \mathbf{H}_{\mathrm{r}k}=\operatorname{Diag}(\mathbf{h}_{\mathrm{r}k})$}.
\end{itemize}

To transform \eqref{eq_passive2} into a convex optimization problem, we first need to deal with the non-convex constraints \eqref{eq_passive2_3}, \eqref{eq_passive2_4}, \eqref{eq_passive2_5} and rank-one constraints \eqref{eq_passive2_9}. Similarly, the first-order Taylor expansion is adopted to acquire the concave lower bounds of left-sides of constraints \eqref{eq_passive2_3}, \eqref{eq_passive2_4}, \eqref{eq_passive2_5} in $q$-th inner loop iteration, denoted as $h\big(\mathbf{Q}_{\mathrm{r}}, \mathbf{Q}_{\mathrm{r}}^{(q)}\big)$, $h_{k, 1}\big(\mathbf{Q}_{\mathrm{t}}, \mathbf{Q}_{\mathrm{t}}^{(q)}\big)$ and $h_{k, 2}\big(\mathbf{Q}_{\mathrm{t}}, \mathbf{Q}_{\mathrm{t}}^{(q)}\big)$.
For the rank-one constraints, we rewrite them as the expressions similar to \eqref{eq_rank-one_rewrite} and add them to the objective function as the penalty terms. Similarly, the linear lower bound of the spectral norm is utilized to replace itself. As a result, the rank-one can be equivalently transformed as
%\begin{align}
%   	\eta_\xi\leq &\operatorname{Tr}(\mathbf{Q}_\xi)-\|\mathbf{Q}_\xi^{(q)}\|_2
%   	-\operatorname{Tr}\big(\mathbf{q}_\xi^{(q)}\big(\mathbf{q}_\xi^{(q)}\big)^H\big(\mathbf{Q}_\xi-\mathbf{Q}_\xi^{(q)}\big)\big) \nonumber\\
%   	=&\widehat{\eta}_\xi\big(\mathbf{Q}_\xi, \mathbf{Q}_\xi^{(q)}\big), \xi\in\{\mathrm{r}, \mathrm{t}\},\label{eq_penal_trans_r}
%   	\eta_\mathrm{t}\leq &\operatorname{Tr}(\mathbf{Q}_\mathrm{t})-\|\mathbf{Q}_\mathrm{t}^{(l)}\|_2
%   	-\operatorname{Tr}\left(\mathbf{q}_\mathrm{t}^{(l)}\left(\mathbf{q}_\mathrm{t}^{(l)}\right)^H\left(\mathbf{Q}_\mathrm{t}-\mathbf{Q}_\mathrm{t}^{(l)}\right)\right) \nonumber\\
%   	=&\widehat{\eta}_\mathrm{t}\left(\mathbf{Q}_\mathrm{t}, \mathbf{Q}_\mathrm{t}^{(l)}\right),\label{eq_penal_trans_t}
%\end{align}
\begin{align}
   	\eta_\xi(\mathbf{Q}_\xi)&\leq \widehat{\eta}_\xi\big(\mathbf{Q}_\xi\big)\triangleq \operatorname{Tr}(\mathbf{Q}_\xi)-\|\mathbf{Q}_\xi^{(q)}\|_2- \nonumber\\
   &\operatorname{Tr}\big(\mathbf{q}_\xi^{(q)}\big(\mathbf{q}_\xi^{(q)}\big)^H\big(\mathbf{Q}_\xi-\mathbf{Q}_\xi^{(q)}\big)\big), ~\xi\in\{\mathrm{r}, \mathrm{t}\},\label{eq_penal_trans_r}
\end{align}
   where $\mathbf{q}_\mathrm{r}^{(q)}$ and $\mathbf{q}_\mathrm{t}^{(q)}$ are the eigenvectors corresponding to the largest eigenvalues of $\mathbf{Q}_\mathrm{r}^{(q)}$ and $\mathbf{Q}_\mathrm{t}^{(q)}$ in $q$-th inner loop iteration. Thus, optimization problem \eqref{eq_passive2} can be re-expressed as
    \begin{subequations}\label{eq_passive3}
   	\begin{align}
   		&\max _{\mathbf{V}}~~ P_1\iota+P_1\kappa+P_0\varpi-\varrho_\mathrm{r}\widehat{\eta}_\mathrm{r}(\mathbf{Q}_\mathrm{r})-\varrho_\mathrm{t}\widehat{\eta}_\mathrm{t}(\mathbf{Q}_\mathrm{t}),\notag \\
   		&~~\text { s.t. } \eqref{eq_passive2_1}, \eqref{eq_passive2_2}, \eqref{eq_passive2_6}-\eqref{eq_passive2_8},\label{eq_passive3_1} \\
   		&\qquad~~~~h\left(\mathbf{Q}_{\mathrm{r}}, \mathbf{Q}_{\mathrm{r}}^{(q)}\right)\geq \iota,\label{eq_passive3_2}\\
   		&\qquad~~~~h_{k, 1}\left(\mathbf{Q}_{\mathrm{r}}, \mathbf{Q}_{\mathrm{r}}^{(q)}\right)\geq \varpi,~\forall{k},\label{eq_passive3_3}\\
   		&\qquad~~~~h_{k, 2}\left(\mathbf{Q}_{\mathrm{r}}, \mathbf{Q}_{\mathrm{r}}^{(q)}\right)\geq\kappa,~\forall{k},\label{eq_passive3_4}
   	 	\end{align}
   \end{subequations}
where $\varrho_\mathrm{r}$ and $\varrho_\mathrm{t}$ denote the penalty coefficients. Thus, SDP optimization problem \eqref{eq_passive3} can be efficiently solved by CVX.
\vspace{-4mm}\subsection{Proposed Optimization Algorithm \& Analysis on Complexity and Convergence}\label{Algorithm}
Algorithm 1 summarizes the proposed iterative algorithm for solving the optimization problem  \eqref{eq_orig_opti} of the STAR-RIS-assisted joint PLS and CC system. The algorithm alternatively solves two subproblems and converges when the objective function gap $v>0$ between two consecutive iterations is below a predefined threshold $\varepsilon$. The penalty violations for active and passive beamforming designs are denoted by $\widehat{v}>0$ and $\tilde{v}>0$, respectively. Note that, the penalty coefficients $\varrho_\mathrm{cs}$, $\varrho_\mathrm{r}$ and $\varrho_\mathrm{t}$ are initialized with small values to prevent the penalty terms from dominating the objective function and leading to inefficient solutions. In addition, $\widehat{\xi}$, $\tilde{\xi}_1$ and $\tilde{\xi}_2$ are the scaling factors for penalty coefficients.
\vspace{-4mm}\begin{center}
	\begin{tabular}{p{8.5cm}}
		\toprule[2pt]
		\small{\textbf{Algorithm 1:}} {\small Proposed Iterative Algorithm for STAR-RIS-assisted joint PLS and CCs  Problem \eqref{eq_orig_opti}}   \\ %solving optimized problem \eqref{eq50}
		\midrule[1pt]
		1: \small{Initialize feasible point $\big(\mathbf{w}_\mathrm{b}^{(0, 0)}, \mathbf{w}_k^{(0, 0)}, \boldsymbol{\Theta}_\mathrm{r}^{(0,0)}, \boldsymbol{\Theta}_\mathrm{t}^{(0,0)}\big)$;} \small{Define} \\ ~~~\small{the tolerance accuracy $\varepsilon$, $\widehat{\varepsilon}$ and $\tilde{\varepsilon}$; Set the outer} \small{iteration index} \\~~~\small{$t$ = 0.}\\
		%~~$i=0$ for  inner loop.  \\
		2: \small{\textbf{While} $v>\varepsilon$  or $t=0$ \textbf{do}}                       \\
		3: \quad \small{Set inner iteration index $i$ = 0; Initialize $\varrho_\mathrm{cs}^{(0)}$}.\\
		4: \quad\small{\textbf{While} $\widehat{v}>\widehat{\varepsilon}$ or $i=0$ \textbf{do}}\\
		5: \qquad \small{Solve the optimization problem \eqref{eq_active4} with the given}\\\qquad~~~\small{$\big(\mathbf{w}_\mathrm{b}^{(t, i)}, \mathbf{w}_k^{(t, i)}, \boldsymbol{\Theta}_\mathrm{r}^{(t,0)}, \boldsymbol{\Theta}_\mathrm{t}^{(t,0)}\big)$ and update $\big(\mathbf{w}_\mathrm{b}^{(t, i+1)},$}\\\qquad~~~\small{$ \mathbf{w}_k^{(t, i+1)}\big)$ with the obtained solutions.}\\
		6: \qquad \small{Calculate $\widehat{v}=\eta_\mathrm{cs}$ based on the acquired solutions;} \small{Update}\\\qquad~~~\small{penalty coefficients $\varrho_\mathrm{cs}=\widehat{\xi} \varrho_\mathrm{cs}$; Let $i=i+1$.} \\
		7: \quad\small{\textbf{end while}}\\
		8: \quad \small{Update $\big(\mathbf{w}_\mathrm{b}^{(t, 0)}, \mathbf{w}_k^{(t, 0)}\big)$ with the $\big(\mathbf{w}_\mathrm{b}^{(t, i)}, \mathbf{w}_k^{(t, i)}\big)$.} \\
		9: \quad Set inner iteration index $q$ = 0; Initialize $\varrho_\mathrm{r}^{(0)}$ and $\varrho_\mathrm{t}^{(0)}$.\\
		10:~~\small{\textbf{While} $\tilde{v}>\tilde{\varepsilon} $ or $q=0$ \textbf{do}}\\
		11:\quad~~\small{Solve the optimization problem \eqref{eq_passive3} with the given $\big(\mathbf{w}_\mathrm{b}^{(t, 0)}$,}\\\qquad ~~\small{$\mathbf{w}_k^{(t, 0)}, \boldsymbol{\Theta}_\mathrm{r}^{(t,q)}, \boldsymbol{\Theta}_\mathrm{t}^{(t,q)}\big)$; Update the $\big(\boldsymbol{\Theta}_\mathrm{r}^{(t, q+1)}, \boldsymbol{\Theta}_\mathrm{t}^{(t, q+1)}\big)$}\\\qquad~~\small{with obtained solutions.}\\
		12:\quad~ \small{Calculate $\tilde{v}=\max\{\eta_\mathrm{r}, \eta_\mathrm{t}\}$ based on the acquired solution;}\\~~~\qquad \small{Update the penalty coefficients $\varrho_\mathrm{r}^{(q+1)}=\tilde{\xi}_1\varrho_\mathrm{r}^{(q)}$,} $\varrho_\mathrm{t}^{(q+1)}=$\\ \quad\qquad \small{$\tilde{\xi}_2\varrho_\mathrm{t}^{(q)}$; Let $q=q+1$.}\\
		13:\quad\small{\textbf{end while}}\\
		14: ~~\small{Update $\big(\mathbf{w}_\mathrm{b}^{(t+1, 0)}, \mathbf{w}_k^{(t+1, 0)}, \boldsymbol{\Theta}_\mathrm{r}^{(t+1,0)}, \boldsymbol{\Theta}_\mathrm{t}^{(t+1,0)}\big)$ with} \\\qquad\small{$\big(\mathbf{w}_\mathrm{b}^{(t, 0)}, \mathbf{w}_k^{(t, 0)}, \boldsymbol{\Theta}_\mathrm{r}^{(t,q)}, \boldsymbol{\Theta}_\mathrm{t}^{(t,q)}\big)$}\\
		15:\quad \small{Calculate the objective value $\overline{R}^{(t+1)}$ and update $v=$}\\\quad\quad~\small{$\left|\overline{R}^{(t+1)}-\overline{R}^{(t)}\right|$ based the obtained solutions; Let $t=t+1$.}\\
		16: \small{\textbf{end while}}      \\
		\bottomrule[2pt]
	\end{tabular}
\end{center}

In terms of the computational complexity of the proposed algorithm, it is mainly dominated by addressing the two standard SDP subproblems\footnote{In the case of solving convex problems, it is presumed that the interior point method is employed, and subsequently, the computational complexity is determined \cite{boyd04}.}.
Specifically, for the joint active beamforming design, the main computed complexity on solving the optimization problem \eqref{eq_active4} can be calculated as $\mathcal{O}\big(\big((K+1)N_\mathrm{t}\big)^{3.5}\big)$.
In the aspect of  joint design the passive beamformer, the calculated complexity comes from the solving of the optimization problem \eqref{eq_passive3}, which is dominated by $\mathcal{O}\left(2M^{3.5}\right)$.
In addition, the bisection search method is utilized to find $\varphi(\epsilon)$ in \eqref{CC_simp} to transform the covert constraint with the computational complexity is $\mathcal{O}\Big(\log_2\big(\frac{s}{\varepsilon_\mathrm{b}}\big)\Big)$, where $s$ and $\varepsilon_\mathrm{b}$ denote the length of the initial search interval and the accuracy tolerance, respectively.
Therefore, the overall computational complexity of the proposed iterative algorithm can be calculated as $\mathcal{O}\Big(\big(\log_2\big(\frac{s}{\varepsilon_\mathrm{b}}\big)+I\big(I_1\Big((K+1)N_\mathrm{t}\Big)^{3.5}+I_2\left(2M^{3.5}\right)\big)\big)\Big)$, where $I$ denotes the total iteration number of the proposed algorithm, $I_1$ and $I_2$ respectively represent the iteration number of the inner loops  for solving two subproblems.
Note that the overall computational complexity  is highly affected by the number of antennas at BS ($N_\mathrm{t}$) and the number of elements equipped at STAR-RIS ($M$).

 Although the alternative strategy is adopted in \mbox{Algorithm 1}, it is easy to verify that the convergence of the proposed iterative algorithm can always  be guaranteed. Note that we can always find a solution not worse than that of the previous iteration in the iterative process. Hence, the objective function value of the optimization problem \eqref{eq_orig_opti} is monotonically non-decreasing w.r.t. the iteration index. Moreover, the convergence of the proposed algorithm will be further proved by simulation results in Section \ref{sec:S5}.

\vspace{-2mm}
\section{Simulation Results}\label{sec:S5}
In this section, the simulation results are presented to validate the effectiveness of the proposed STAR-RIS-aided joint PLS and CCs scheme. In particular, we assume that the mmWave communication system assisted by STAR-RIS operates at 28 GHz with bandwidth 251.1886 MHz. Hence, the noise power can be calculated as $\sigma_\mathrm{b}^2=-90$ dBm and $\sigma_{k}^2=-90$ dBm.
In addition, we consider that the simulated system has $K=3$ security users and set the QoS minimum rates as $R_\mathrm{b}^*=0.5$, $R_\mathrm{s0}^*=0.6$ and $R_\mathrm{s1}^*=0.6$.
For the large-scale path loss values in \eqref{HBR} and \eqref{h_r}, the theoretical free-space distance-dependent path-loss model \cite{feng2022joint} is leveraged, which is given by $l_{\varpi}=-30-22\log{d_{\varpi}}$ dB, $\varpi\in\{\mathrm{BR}, \mathrm{rb}, \mathrm{r}k\}$ for $k\in\mathcal{K}$.
 The distances are set as $d_\mathrm{BR}=40$ m, $d_\mathrm{rb}=15$ m and $d_{\mathrm{r}k}=15$ m. Moreover, the tolerance accuracy $\varepsilon$, $\widehat{\varepsilon}$ and $\tilde{\varepsilon}$ in the proposed iterative algorithm are set as $10^{-4}$, $10^{-6}$ and $10^{-6}$, respectively. To highlight the potential of STAR-RIS in jointly implementing the PLS and CCs, a baseline scheme is proposed where two adjacent conventional RISs with $\frac{M}{2}$ elements are adopted to replace the STAR-RIS.
\begin{figure}[ht]
	\centering
	\includegraphics[scale=0.4]{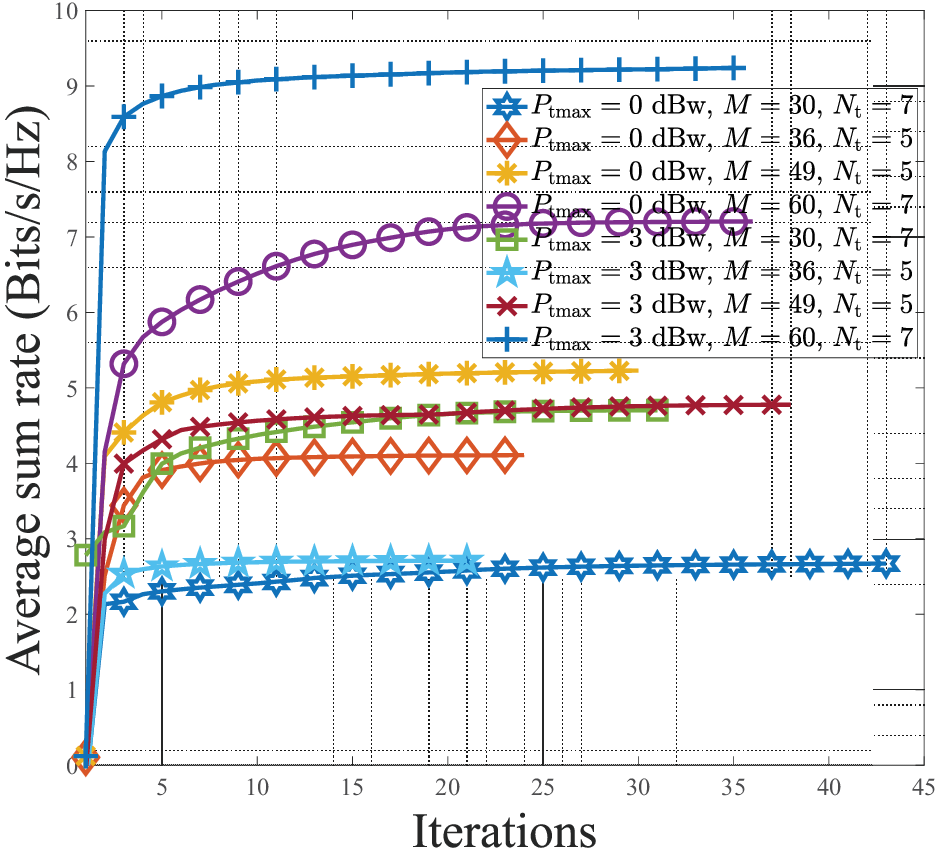}\\
	\caption{Average sum rate versus the iterations with $\epsilon=0.1$ and $P_1=0.5$, and different $P_\mathrm{tmax}$, $M$ and $N_\mathrm{t}$.}\label{fig:iterations}
\end{figure}

The convergence curves of the proposed iterative algorithm are depicted in Fig. \ref{fig:iterations}, taking into account of various maximum transmit power at the BS, as well as the number of elements and antennas equipped at the STAR-RIS and the BS. Specifically, we conducts evaluation of convergence using eight diverse cases for the proposed method. The presented results indicate that the obtained sum rates exhibit a monotonically non-decreasing behavior versus the number of iterations. In addition, the proposed algorithm consistently achieves rapid convergence to a stable value within a few iterations. Hence, the efficiency of the proposed algorithm can be validated.
\begin{figure}[ht]
	\centering
	\includegraphics[scale=0.4]{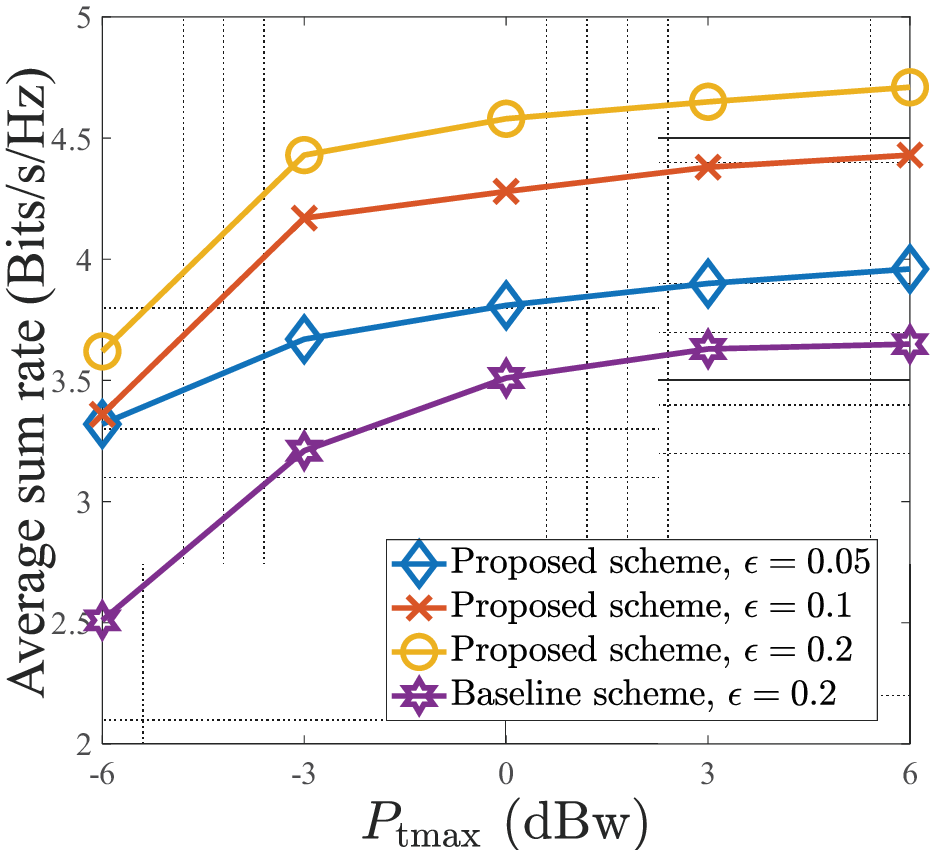}\\
	\caption{Average sum rate versus the maximum transmit power $P_\mathrm{tmax}$ at BS with $M=30$, $N_\mathrm{t}=7$, $P_1=0.5$, and different covert requirements $\epsilon$.}\label{fig:Pvsrate}
\end{figure}

Fig. \ref{fig:Pvsrate} presents the variation curves of the average sum rates versus the maximal transmit power $P_\mathrm{tmax}$ with different covert requirements  $\epsilon$, in comparison with the baseline utilizing the traditional RIS. It can be observed that the average sum rates gradually increase w.r.t. $P_\mathrm{tmax}$ in all cases, indicating that there exists a positive correlation between the average sum rates and $P_\mathrm{tmax}$.
However, the speeds of increase diminish with the growth of the maximum transmit power. Additionally, a relaxed covert requirement contributes to breaking through the performance bottleneck constrained by other system indicators.
It is obvious that the proposed scheme exhibits significant performance benefits in jointly implementing the PLS and CCs in comparison to the baseline scheme. The proposed scheme can achieve better performance even if  it is operated at a tighter covert requirement (i.e., $\epsilon=0.05$).
\begin{figure}[ht]
	\centering
	\includegraphics[scale=0.38]{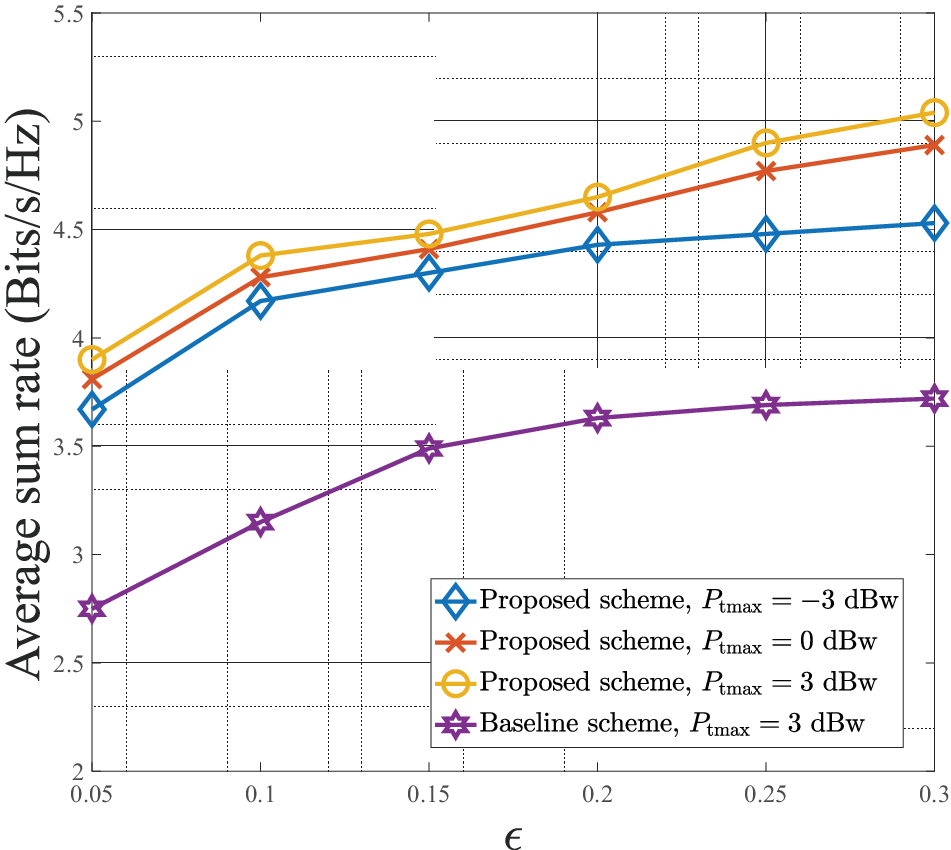}\\
	\caption{Average sum rate versus the covert requirement $\epsilon$ with $M=30$, $N_\mathrm{t}=7$, $P_1=0.5$, and different $P_\mathrm{tmax}$.}\label{fig:epsilonvsrate}
\end{figure}

Next, we investigate the influence of the covert requirements $\epsilon$ on the average sum rate considering  different $P_\mathrm{tmax}$, as shown in Fig. \ref{fig:epsilonvsrate}. According to the given results, we can find that the average sum rates increase progressively versus $\epsilon$ in all scenarios due to the fact that the covert requirement becomes more relaxed. To acquire an apparent comparison, $P_\mathrm{tmax}=3$ dBw is selected to operate the baseline scheme. Despite this, the achieved performance gain of the baseline falls significantly short to that of the proposed scheme, even if the proposed scheme is operated at a much lower maximum transmit power of $P_\mathrm{tmax}=-3$ dBw. Therefore, the STAR-RIS-aided scheme shows a great advantage in enhancing the system performance as compared to the conventional RIS.
\begin{figure}[ht]
	\centering
	\includegraphics[scale=0.4]{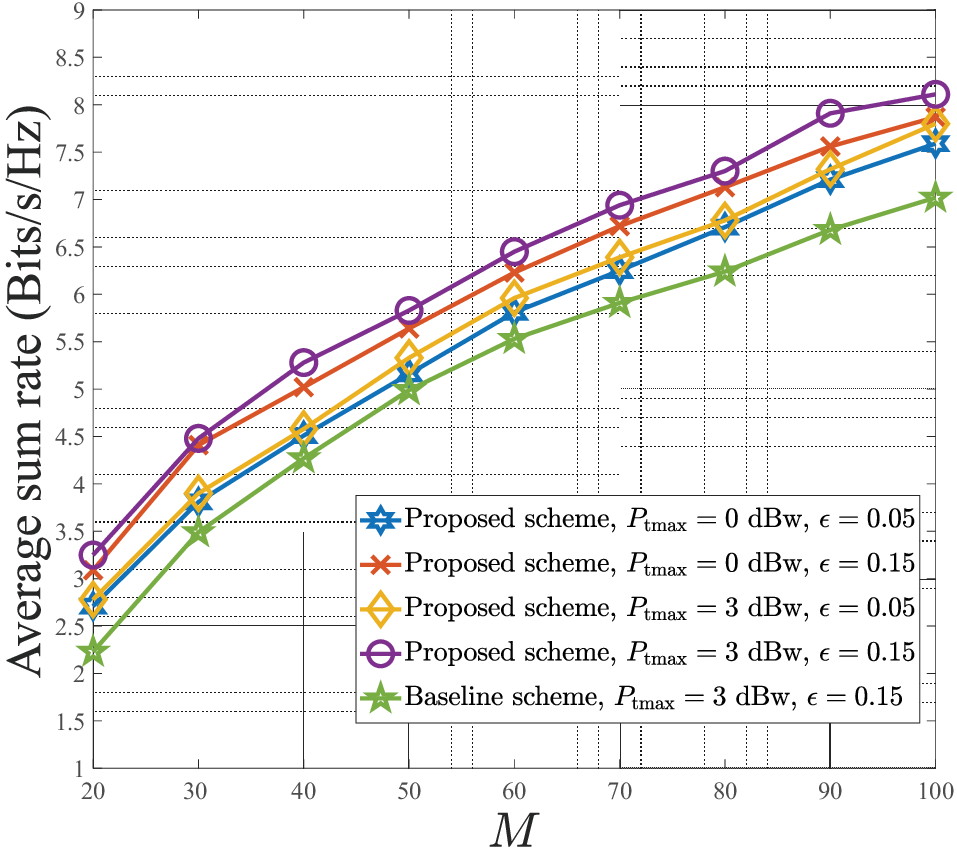}\\
	\caption{Average sum rate versus the number of elements equipped at STAR-RIS with $N_\mathrm{t}=7$, $P_1=0.5$, and different maximum transmit power $P_\mathrm{tmax}$ and covert requirements $\epsilon$.}\label{fig:Mvsrate}
\end{figure}

In Fig. \ref{fig:Mvsrate}, the performance trends of the average sum rate w.r.t. the number of elements at STAR-RIS ($M$)  are presented, taking into account of various $P_\mathrm{tmax}$ and covert requirements $\epsilon$.
In particular, it is discernible that the average sum rates exhibit ascending trends with the increased $M$, which is due to the fact that more elements can provide a higher degree of freedom to augment performance gains.
Besides, we also find that the relaxed covert demands may offer more potential to break through the performance limitation imposed by the system settings than the incremental maximal transmit power by respectively comparing the simulation results with the same $P_\mathrm{tmax}$ and $\epsilon$. Similarly, the most relaxed condition (i.e., $P_\mathrm{tmax}=3$ dBw, $\epsilon=0.15$) is adopted to implement the baseline scheme, however, the acquired performance is still worse than the proposed scheme under the strictest condition (i.e., $P_\mathrm{tmax}=0$ dBw, $\epsilon=0.05$).
\begin{figure}[ht]
	\centering
	\includegraphics[scale=0.4]{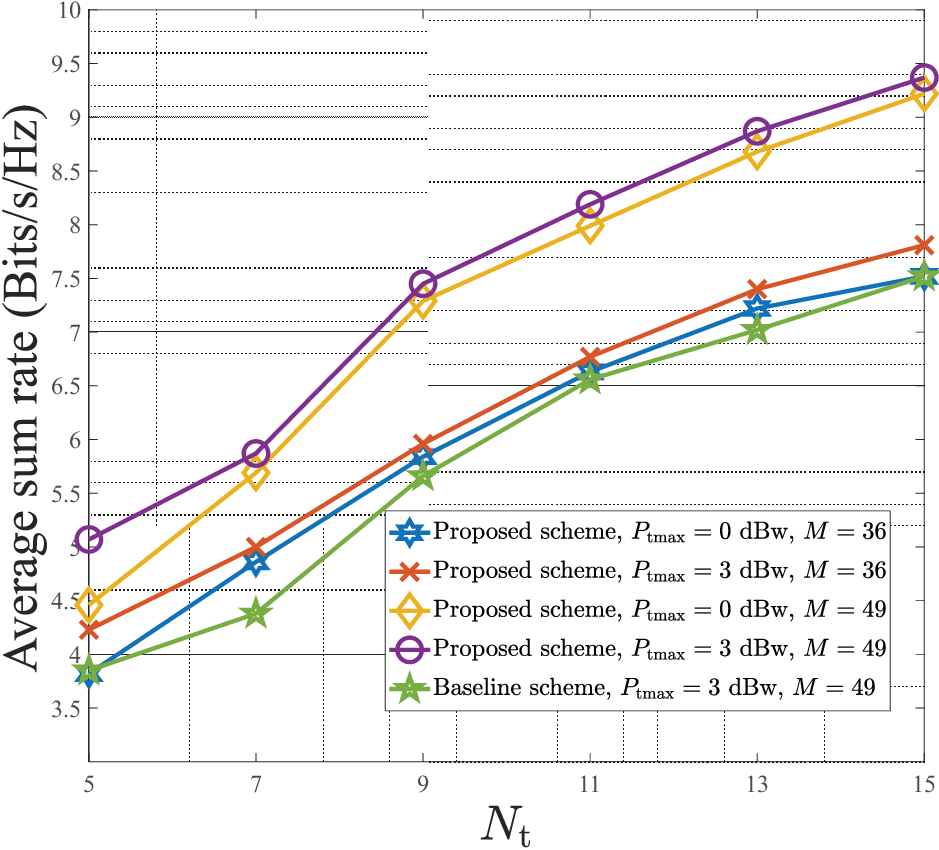}\\
	\caption{Average sum rate versus the number of antennas at BS with $\epsilon=0.1$, $P_1=0.5$, and different maximal transmit power $P_\mathrm{tmax}$ and $M$.}\label{fig:Ntvsrate}
\end{figure}	`

We explore the impact of the number of antennas installed at the BS ($N_\mathrm{t}$) on the system performance in Fig. \ref{fig:Ntvsrate} with different $P_\mathrm{tmax}$ and $M$. Specifically, a comparable performance trend can still be noted, wherein the average sum rate gradually raises as the number of antennas at the BS is augmented. In addition, we can observe that  increasing the number of elements at the STAR-RIS from $M=36$ to $M=49$ can achieve much more performance improvement than enlarging $P_\mathrm{tmax}$ from 0dBw to 3dBw. Although we choose  $P_\mathrm{tmax}=3$ dBw and $M=49$ to operate the baseline scheme, its presented performance gain is still below the proposed scheme under the most rigorous  condition (i.e., $P_\mathrm{tmax}=0$ dBw, $M=36$), which further indicates the superiority of STAR-RIS in ensuring the performance of joint PLS and CCs.

\begin{figure}[h]
	\centering
	\includegraphics[scale=0.33]{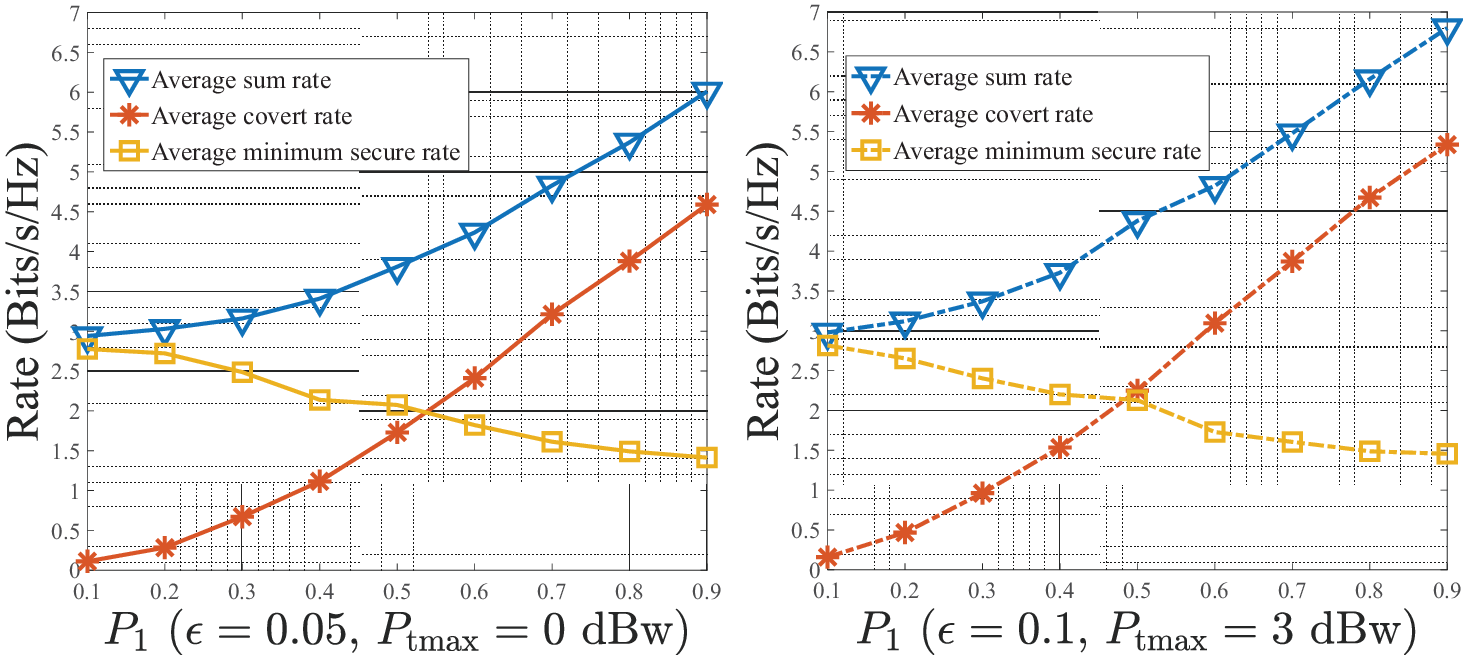}\\
	\caption{Average sum rate, average covert rate and average minimum secure rate versus the covert transmission probability $P_1$ with $M=30$, $N_\mathrm{t}=7
		$, and different $P_\mathrm{tmax}$ and $\epsilon$.}\label{fig:p1tvsrate}
\end{figure}

Lastly, we investigate the influence of the probability for CCs to Bob, i.e., $P_1$, on the average sum rate, the average covert rate ( $P_1 R_\mathrm{b}^\mathrm{c}$), and the average minimum secure rate ($P_ 0\min_{k}\widehat{R}^k_\mathrm{sl, 0}+P_ 1\min_{k}\widehat{R}^k_\mathrm{sl, 1}$), in a time slot, as presented in Fig. \ref{fig:p1tvsrate}. According to the simulated results, it can be concluded that an increase in the value of $P_1$ from 0.1 to 0.9 leads to an upward trend in both the average sum rate and the average covert rate. Conversely, the average minimum secure rate displays a downward trend, suggesting that the improvement in the average sum rate versus $P_1$ is mainly attributed to the average covert rate. The rationale behind this phenomenon is due to the fact that the  energy splitting protocol is adopted at the STAR-RIS. As a result,  more amplitude energy will be allocated to the reflected coefficients while less to the transmitted coefficients as the CCs probability $P_1$ increases, which will lead to an increase of the covert rate while a decrease of the secure rate.
%the reflected coefficients provided by the STAR-RIS is completely leveraged to improve the covert rate, while the transmitted coefficients is designed by simultaneously considering $K$ users' secure rate in the considered simulation system of this paper.

\vspace{-2mm}\section{Conclusion}\label{sec:S6}
In this paper, we initially investigate the STAR-RIS enhanced joint PLS and CCs for mmWave systems. In particular, the analytical derivations of the minimum DEP and the lower bound of the secure rates are obtained by considering the practical assumptions, where only the statistical CSI between STAR-RIS and the wardens is accessible at the BS.
An optimization problem is constructed that focuses on maximizing the average sum rate between the  covert rate and the minimum secure rate, while also ensuring the covert constraint and QoS constraints. In order to effectively solve this non-convex optimization problem with strong coupling variables, an alternative algorithm based on the SDR method is proposed. Numerical results demonstrate the performance gains of the  proposed STAR-RIS-assisted scheme in comparison with the benchmark scheme adopting the traditional RIS, which further indicates that the STAR-RIS exhibits more benefits in the implementation of the joint PLS and CCs.
\appendices
\vspace{-4mm}\section{Proof of Theorem 1}\label{Apd_A}
In this section, we derive the FA probability and MD probability in detail. In particular, we first give the expression of $P_\mathrm{FA}$ based on the definition of FA probability, which is expressed as
\begin{align}
	P_\mathrm{FA}=&\operatorname{Pr}\left(P_\mathrm{w}>\tau_{dt}|\mathcal{H}_0\right)\notag\\
	=&\operatorname{Pr}\left(\sum_{k=1}^K\left|\mathbf{h}_\mathrm{rw}^H\boldsymbol{\Theta}_\mathrm{r}\mathbf{H}_\mathrm{BR}\mathbf{w}_k\right|^2+\sigma^2_\mathrm{w}>\tau_\mathrm{dt}\right).
%	=&\begin{cases}1,&\tau_\mathrm{dt}\leq\sigma_\mathrm{w}^2,\\
%		\int_{\tau_\mathrm{dt}-\sigma_\mathrm{w}^2}^{+\infty}\frac{e^{-\frac{x}{\lambda_0}}}{\lambda_0}dx,&otherswhile,
%	\end{cases}.
%	=&\begin{cases}1,&\tau_\mathrm{dt}\leq\sigma_\mathrm{w}^2,\\
%		e^{-\frac{\tau_\mathrm{dt}-\sigma_\mathrm{w}^2}{\lambda_0}},&otherswhile,
%	\end{cases}
\end{align}
Considering only the statistical CSI of $\mathbf{H}_\mathrm{BR}$ is available at Willie, thus letting $\chi_k=\left|\mathbf{h}_\mathrm{rw}^H\boldsymbol{\Theta}_\mathrm{r}\mathbf{H}_\mathrm{BR}\mathbf{w}_k\right|^2$, we have
\begin{align}
	\chi_k=&\left|\operatorname{Tr}(\mathbf{h}_\mathrm{rw}^H\boldsymbol{\Theta}_\mathrm{r}\mathbf{H}_\mathrm{BR}\mathbf{w}_k)|\right.^2
	=\left|\operatorname{Tr}(\mathbf{H}_\mathrm{BR}\mathbf{w}_k\mathbf{h}_\mathrm{rw}^H\boldsymbol{\Theta}_\mathrm{r})|\right.^2\notag\\
	\stackrel{(a)}{=}&\left|\operatorname{vec}(\mathbf{H}_\mathrm{BR})^T\operatorname{vec}\left((\mathbf{w}_k\mathbf{h}_\mathrm{rw}^H\boldsymbol{\Theta}_\mathrm{r})^T\right)\right|^2,
\end{align}
where $(a)$ is due to \cite[(eq.1.11.12)]{zhang2017matrix}. Besides, it is easy to verify that $\operatorname{vec}(\mathbf{H}_\mathrm{BR})\sim \mathcal{C N}(\mathbf{0}, \frac{N_\mathrm{t}M\rho_\mathrm{BR}}{L}\boldsymbol{\Phi}^H\boldsymbol{\Phi})$. Hence, we can observe that $\chi_k$ is an exponential random variable whose probability density function (PDF) is $f_{\chi_k}(x)=\frac{e^{\frac{-x}{\lambda_k}}}{\lambda_k}$, where $\lambda_k=\frac{N_\mathrm{t}M\rho_\mathrm{BR}}{L}\left\|\boldsymbol{\Phi}\operatorname{vec}\left((\mathbf{w}_\mathrm{b}\mathbf{h}_
\mathrm{rw}^H\boldsymbol{\Theta}_\mathrm{r})^T\right)\right\|^2_2$.

According to the above analysis, the analytic expressions of $P_\mathrm{FA}$ and $P_\mathrm{MD}$ can be derived as
\begin{align}
	P_\mathrm{FA}=&\begin{cases}1,\notag\\
		\int_{\tau_\mathrm{dt}-\sigma_\mathrm{w}^2}^{+\infty}\frac{e^{-\frac{x}{\lambda_0}}}{\lambda_0}dx,
	\end{cases}
	=\begin{cases}1,&\tau_\mathrm{dt}\leq\sigma_\mathrm{w}^2,\\
	e^{-\frac{\tau_\mathrm{dt}-\sigma_\mathrm{w}^2}{\lambda_0}},&\mathrm{otherwise},
	\end{cases}\\
P_\mathrm{MD}=&\operatorname{Pr}\Big(\left|\mathbf{h}_\mathrm{rw}^H
\boldsymbol{\Theta}_\mathrm{r}\mathbf{H}_\mathrm{BR}\mathbf{w}_\mathrm{b}\right|^2
+\sum_{k=1}^K\left|\mathbf{h}_\mathrm{rw}^H\boldsymbol{\Theta}_\mathrm{r}\mathbf{H}_\mathrm{BR}\mathbf{w}_k\right|^2\notag\\
&+\sigma^2_\mathrm{w}<\tau_\mathrm{dt}\Big)\notag\\
=&\begin{cases}0,\notag\\
	\int_{0}^{\tau_\mathrm{dt}-\sigma_\mathrm{w}^2}\frac{e^{-\frac{x}{\lambda_1}}}{\lambda_1}dx,
\end{cases}
=\begin{cases}0,&\tau_\mathrm{dt}\leq\sigma_\mathrm{w}^2,\\
	1-e^{-\frac{\tau_\mathrm{dt}-\sigma_\mathrm{w}^2}{\lambda_1}},&\mathrm{otherwise}.
\end{cases}
\end{align}
\vspace{-4mm}\section{Proof of Theorem 2}\label{Apd_B}
The asymptotic results of $\lambda_0$ and $\lambda_1$ leveraging the large system analytic technique are derived in this section. Specifically, we first equivalently rewrite $\psi=\left\|\boldsymbol{\Phi}\operatorname{vec}\left((\mathbf{w}_k\mathbf{h}_\mathrm{rw}^H\boldsymbol{\Theta}_\mathrm{r})^T\right)\right\|^2_2$ as
\begin{align}\label{eq_psi_rewrite}
	\psi=&\operatorname{vec}\left((\mathbf{w}_k\mathbf{h}_
	\mathrm{rw}^H\boldsymbol{\Theta}_\mathrm{r})^T\right)^H\boldsymbol{\Phi}^H\boldsymbol{\Phi}
	\operatorname{vec}\left((\mathbf{w}_k\mathbf{h}_\mathrm{rw}^H\boldsymbol{\Theta}_\mathrm{r})^T\right)
	\notag\\=&\left\|\boldsymbol{\Phi}\operatorname{vec}\left((\mathbf{w}_k\mathbf{h}_\mathrm{rw}^H\boldsymbol{\Theta}_\mathrm{r})^T\right) \right\|^2_2
	\stackrel{(a)}{=}\left\| \boldsymbol{\Phi}\left(\mathbf{w}_k\otimes\boldsymbol{\Theta}_\mathrm{r}\right)\mathbf{h}_\mathrm{rw}^* \right\|^2_2\notag\\
	=&\frac{M\rho_\mathrm{rw}}{P}\left\|\boldsymbol{\Phi}\left(\mathbf{w}_k\otimes\boldsymbol{\Theta}_\mathrm{r}\right)\boldsymbol{\Omega}_\mathrm{rw}^H\mathbf{g}^* \right\|^2_2,
\end{align}
where $\boldsymbol{\Omega}_\mathrm{rw}=\left[\mathbf{a}_\mathrm{R}\left(\phi_1^\mathrm{rw}, \theta_1^\mathrm{rw}\right), \cdots,
\mathbf{a}_\mathrm{R}\left(\phi_{P}^\mathrm{rw}, \theta_{P}^\mathrm{rw}\right)\right]^H$, $\mathbf{g}=[g^\mathrm{rw}_1, \cdots, g^\mathrm{rw}_P]^T$,
and $(a)$ is based on the result from \cite[(eq. 1.11.18)]{zhang2017matrix}. Then, by applying the large system analytical method in \cite{evans00} on \eqref{eq_psi_rewrite}, we can further  obtain \eqref{eq_ large_system} given below
{\small{
 \begin{align}\label{eq_ large_system}
 		&\lim_{M\rightarrow\infty}\frac{\left\| \boldsymbol{\Phi}\big(\mathbf{w}_k\otimes\boldsymbol{\Theta}_\mathrm{r}\big) \boldsymbol{\Omega}_\mathrm{rw}^H\mathbf{g}^*\right\|^2_2}{M}\notag\\
 		&=\lim_{M\rightarrow\infty}\frac{\operatorname{Tr}\big(\mathbf{g}^T\boldsymbol{\Omega}_\mathrm{rw}\left(\mathbf{w}_k\otimes\boldsymbol{\Theta}_\mathrm{r}\right)^H\boldsymbol{\Phi}^H
 			\boldsymbol{\Phi}(\mathbf{w}_k\otimes\boldsymbol{\Theta}_\mathrm{r})\boldsymbol{\Omega}_\mathrm{rw}^H\mathbf{g}^*\big)}{M}\notag\\
 		&\stackrel{(a)}{\rightarrow}\frac{\operatorname{Tr}\big(\boldsymbol{\Omega}_\mathrm{rw}^H\boldsymbol{\Omega}_\mathrm{rw}\big(\mathbf{w}_k\otimes\boldsymbol{\Theta}_\mathrm{r}\big)^H\boldsymbol{\Phi}^H
 			\boldsymbol{\Phi}(\mathbf{w}_k\otimes\boldsymbol{\Theta}_\mathrm{r})\big)}{M}\notag\\
 		&\stackrel{(b)}{=}\frac{\operatorname{Tr}\Big(\sum\limits_{l=1}^{L}\big(\mathbf{w}^H_k\boldsymbol{\Psi}^l_\mathrm{BR}\mathbf{w}_k\big)
 			\otimes\big(\boldsymbol{\Omega}_\mathrm{rw}^H\boldsymbol{\Omega}_\mathrm{rw}\boldsymbol{\Theta}^H_\mathrm{r}\widehat{\boldsymbol{\Psi}}^l_\mathrm{BR}\boldsymbol{\Theta}_\mathrm{r}\big)\Big)}{M}\notag\\
 		&\stackrel{(c)}{=}\frac{\sum\limits_{l=1}^{L}\big(\mathbf{w}^H_k\boldsymbol{\Psi}^l_\mathrm{BR}\mathbf{w}_k\big)\operatorname{Tr}\big(\boldsymbol{\Omega}_\mathrm{rw}^H\boldsymbol{\Omega}_\mathrm{rw}
 			\boldsymbol{\Theta}^H_\mathrm{r}\widehat{\boldsymbol{\Psi}}^l_\mathrm{BR}\boldsymbol{\Theta}_\mathrm{r}\big)}{M}\notag\\
% 		&\stackrel{(d)}{=}\frac{1}{M}\sum\limits_{l=1}^{L}\big(\mathbf{w}^H_k\boldsymbol{\Psi}^l_\mathrm{BR}\mathbf{w}_k\big)
% 			\big(\operatorname{vec}^T\big(\boldsymbol{\Theta}_\mathrm{r}^T\big)\big(
% 			\big(\widehat{\boldsymbol{\Psi}}^l_\mathrm{BR}\big)^T\notag\\
% 			&\otimes\big(\boldsymbol{\Omega}_\mathrm{rw}^H\boldsymbol{\Omega}_\mathrm{rw}\big)\big)\operatorname{vec}\big(\boldsymbol{\Theta}_\mathrm{r}^H\big)\big)\notag\\
 		&=\frac{\sum\limits_{l=1}^{L}\big(\mathbf{w}^H_k\boldsymbol{\Psi}^l_\mathrm{BR}\mathbf{w}_k\big)
 			\big(\boldsymbol{\vartheta}_\mathrm{r}^T\boldsymbol{\Xi}^T\big(\big(\widehat{\boldsymbol{\Psi}}^l_\mathrm{BR}\big)^T\otimes\big(\boldsymbol{\Omega}_\mathrm{rw}^H\boldsymbol{\Omega}_\mathrm{rw}\big)\big)\boldsymbol{\Xi}\boldsymbol{\vartheta}_\mathrm{r}^*\big)}{M},
 \end{align}
}}
%where $\boldsymbol{\vartheta}_\mathrm{r}=\operatorname{diag}(\boldsymbol{\Theta}_\mathrm{r})$, $	\boldsymbol{\Phi}^H\boldsymbol{\Phi}=\sum_{l=1}^{L}\boldsymbol{\Psi}^l_\mathrm{BR}\otimes\widehat{\boldsymbol{\Psi}}^l_\mathrm{BR}$, $\boldsymbol{\Psi}^l_\mathrm{BR}=\mathbf{a}_\mathrm{B}\left(\gamma_l^{\mathrm{BR}}\right)\mathbf{a}_\mathrm{B}^H\left(\gamma_l^{\mathrm{BR}}\right)$, $
%\widehat{\boldsymbol{\Psi}}^l_\mathrm{BR}=\mathbf{a}_\mathrm{R}\left(\phi_l^\mathrm{BR}, \theta_l^\mathrm{BR}\right)\mathbf{a}_\mathrm{R}^H\left(\phi_l^\mathrm{BR}, \theta_l^\mathrm{BR}\right)$, $ \boldsymbol{\Xi}=\big[[\mathbf{e}_{1}, \mathbf{0}_{M\times(M-1)}];$ $[\mathbf{0}_{M\times1}, \mathbf{e}_{2}, \mathbf{0}_{M\times(M-2)}]; \cdots;[\mathbf{0}_{M\times(M-1)}, \mathbf{e}_{M} ]\big]$;
The convergence $(a)$ is from the corollary in  \cite[Corollary 1]{evans00}; steps $(b)$ and $(c)$ are because of \cite[(eq. 1.10.15)]{zhang2017matrix} and \cite[(eq. 1.10.11)]{zhang2017matrix}, respectively. On the basis of the derived result in \eqref{eq_ large_system}, the asymptotic results of $\lambda_0$ and $\lambda_1$ can be obtained as in \eqref{eq_lambda_0a} and \eqref{eq_lambda_1a}.

\vspace{-4mm}
\ifCLASSOPTIONcaptionsoff %\emph{a}
  \newpage
\fi
\bibliographystyle{IEEEtran}
\bibliography{PSL-CC}

\begin{thebibliography}{10}
\providecommand{\url}[1]{#1}
\csname url@samestyle\endcsname
\providecommand{\newblock}{\relax}
\providecommand{\bibinfo}[2]{#2}
\providecommand{\BIBentrySTDinterwordspacing}{\spaceskip=0pt\relax}
\providecommand{\BIBentryALTinterwordstretchfactor}{4}
\providecommand{\BIBentryALTinterwordspacing}{\spaceskip=\fontdimen2\font plus
\BIBentryALTinterwordstretchfactor\fontdimen3\font minus
  \fontdimen4\font\relax}
\providecommand{\BIBforeignlanguage}[2]{{%
\expandafter\ifx\csname l@#1\endcsname\relax
\typeout{** WARNING: IEEEtran.bst: No hyphenation pattern has been}%
\typeout{** loaded for the language `#1'. Using the pattern for}%
\typeout{** the default language instead.}%
\else
\language=\csname l@#1\endcsname
\fi
#2}}
\providecommand{\BIBdecl}{\relax}
\BIBdecl

\bibitem{Wyner1975wire}
A.~D. Wyner, ``The wire-tap channel,'' \emph{The Bell Syst. Tech. J.}, vol.~54,
  no.~8, pp. 1355--1387, 1975.

\bibitem{yang2012transmit}
N.~Yang, P.~L. Yeoh, M.~Elkashlan, R.~Schober, and I.~B. Collings, ``Transmit
  antenna selection for security enhancement in {MIMO} wiretap channels,''
  \emph{IEEE Trans. Commun.}, vol.~61, no.~1, pp. 144--154, 2012.

\bibitem{lv2015secrecy}
T.~Lv, H.~Gao, and S.~Yang, ``Secrecy transmit beamforming for heterogeneous
  networks,'' \emph{IEEE J. Sel. Areas Commun.}, vol.~33, no.~6, pp.
  1154--1170, 2015.

\bibitem{zhao2017artificial}
N.~Zhao, Y.~Cao, F.~R. Yu, Y.~Chen, M.~Jin, and V.~C. Leung, ``Artificial noise
  assisted secure interference networks with wireless power transfer,''
  \emph{IEEE Trans. Veh. Technol.}, vol.~67, no.~2, pp. 1087--1098, 2017.

\bibitem{hu2016secrecy}
X.~Hu, P.~Mu, B.~Wang, and Z.~Li, ``On the secrecy rate maximization with
  uncoordinated cooperative jamming by single-antenna helpers,'' \emph{IEEE
  Trans. Veh. Technol.}, vol.~66, no.~5, pp. 4457--4462, 2016.

\bibitem{zheng2022physical}
T.-X. Zheng, X.~Chen, C.~Wang, K.-K. Wong, and J.~Yuan, ``Physical layer
  security in large-scale random multiple access wireless sensor networks: a
  stochastic geometry approach,'' \emph{IEEE Trans. Commun.}, vol.~70, no.~6,
  pp. 4038--4051, 2022.

\bibitem{zheng21}
T.-X. Zheng, Z.~Yang, C.~Wang, Z.~Li, J.~Yuan, and X.~Guan, ``Wireless covert
  communications aided by distributed cooperative jamming over slow fading
  channels,'' \emph{IEEE Trans. Wireless Commun.}, vol.~20, no.~11, pp.
  7026--7039, 2021.

\bibitem{yan2019low}
S.~Yan, X.~Zhou, J.~Hu, and S.~V. Hanly, ``Low probability of detection
  communication: Opportunities and challenges,'' \emph{IEEE Wireless Commun.},
  vol.~26, no.~5, pp. 19--25, 2019.

\bibitem{bash2013limits}
B.~A. Bash, D.~Goeckel, and D.~Towsley, ``Limits of reliable communication with
  low probability of detection on {AWGN} channels,'' \emph{IEEE J. Sel. Areas
  Commun.}, vol.~31, no.~9, pp. 1921--1930, 2013.

\bibitem{goeckel2015covert}
D.~Goeckel, B.~Bash, S.~Guha, and D.~Towsley, ``Covert communications when the
  warden does not know the background noise power,'' \emph{IEEE Commun. Lett.},
  vol.~20, no.~2, pp. 236--239, 2015.

\bibitem{he2017covert}
B.~He, S.~Yan, X.~Zhou, and V.~K. Lau, ``On covert communication with noise
  uncertainty,'' \emph{IEEE Commun. Lett.}, vol.~21, no.~4, pp. 941--944, 2017.

\bibitem{wang2018covert}
J.~Wang, W.~Tang, Q.~Zhu, X.~Li, H.~Rao, and S.~Li, ``Covert communication with
  the help of relay and channel uncertainty,'' \emph{IEEE Wireless Commun.
  Lett.}, vol.~8, no.~1, pp. 317--320, 2018.

\bibitem{chen2021multi}
X.~Chen, W.~Sun, C.~Xing, N.~Zhao, Y.~Chen, F.~R. Yu, and A.~Nallanathan,
  ``Multi-antenna covert communication via full-duplex jamming against a warden
  with uncertain locations,'' \emph{IEEE Trans. Wireless Commun.}, vol.~20,
  no.~8, pp. 5467--5480, 2021.

\bibitem{li2020optimal}
K.~Li, P.~A. Kelly, and D.~Goeckel, ``Optimal power adaptation in covert
  communication with an uninformed jammer,'' \emph{IEEE Trans. Wireless
  Commun.}, vol.~19, no.~5, pp. 3463--3473, 2020.

\bibitem{zheng2019multi}
T.-X. Zheng, H.-M. Wang, D.~W.~K. Ng, and J.~Yuan, ``Multi-antenna covert
  communications in random wireless networks,'' \emph{IEEE Trans. Wireless
  Commun.}, vol.~18, no.~3, pp. 1974--1987, 2019.

\bibitem{shahzad2019covert}
K.~Shahzad, X.~Zhou, and S.~Yan, ``Covert wireless communication in presence of
  a multi-antenna adversary and delay constraints,'' \emph{IEEE Trans. Veh.
  Technol.}, vol.~68, no.~12, pp. 12\,432--12\,436, 2019.

\bibitem{cui2019secure}
M.~Cui, G.~Zhang, and R.~Zhang, ``Secure wireless communication via intelligent
  reflecting surface,'' \emph{IEEE Wireless Commun. Lett.}, vol.~8, no.~5, pp.
  1410--1414, 2019.

\bibitem{dong2021double}
L.~Dong, H.-M. Wang, J.~Bai, and H.~Xiao, ``Double intelligent reflecting
  surface for secure transmission with inter-surface signal reflection,''
  \emph{IEEE Trans. Veh. Technol.}, vol.~70, no.~3, pp. 2912--2916, 2021.

\bibitem{lu2020intelligent}
X.~Lu, E.~Hossain, T.~Shafique, S.~Feng, H.~Jiang, and D.~Niyato, ``Intelligent
  reflecting surface enabled covert communications in wireless networks,''
  \emph{IEEE Netw.}, vol.~34, no.~5, pp. 148--155, 2020.

\bibitem{zhou2021intelligent}
X.~Zhou, S.~Yan, Q.~Wu, F.~Shu, and D.~W.~K. Ng, ``Intelligent reflecting
  surface {(IRS)}-aided covert wireless communications with delay constraint,''
  \emph{IEEE Trans. Wireless Commun.}, vol.~21, no.~1, pp. 532--547, 2021.

\bibitem{X.Hu21}
X.~Hu, C.~Masouros, and K.-K. Wong, ``Reconfigurable intelligent surface aided
  mobile edge computing: {F}rom optimization-based to location-only
  learning-based solutions,'' \emph{IEEE Trans. Commun.}, vol.~69, no.~6, pp.
  3709--3725, 2021.

\bibitem{liu2021star}
Y.~Liu, X.~Mu, J.~Xu, R.~Schober, Y.~Hao, H.~V. Poor, and L.~Hanzo, ``{STAR}:
  Simultaneous transmission and reflection for 360$^\circ$ coverage by
  intelligent surfaces,'' \emph{IEEE Wireless Commun.}, vol.~28, no.~6, pp.
  102--109, 2021.

\bibitem{mu2021simultaneously}
X.~Mu, Y.~Liu, L.~Guo, J.~Lin, and R.~Schober, ``Simultaneously transmitting
  and reflecting {(STAR) RIS} aided wireless communications,'' \emph{IEEE
  Trans. Wireless Commun.}, vol.~21, no.~5, pp. 3083--3098, 2021.

\bibitem{han2022artificial}
Y.~Han, N.~Li, Y.~Liu, T.~Zhang, and X.~Tao, ``Artificial noise aided secure
  {NOMA} communications in {STAR-RIS} networks,'' \emph{IEEE Wireless Commun.
  Lett.}, vol.~11, no.~6, pp. 1191--1195, 2022.

\bibitem{zhang2022secrecy}
Z.~Zhang, J.~Chen, Y.~Liu, Q.~Wu, B.~He, and L.~Yang, ``On the secrecy design
  of {STAR-RIS} assisted uplink {NOMA} networks,'' \emph{IEEE Trans. Wireless
  Commun.}, vol.~21, no.~12, pp. 11\,207--11\,221, 2022.

\bibitem{xiao2023simultaneously}
H.~Xiao, X.~Hu, P.~Mu, W.~Wang, T.-X. Zheng, K.-K. Wong, and K.~Yang,
  ``Simultaneously transmitting and reflecting {RIS (STAR-RIS)} assisted
  multi-antenna covert communications: Analysis and optimization,'' \emph{arXiv
  preprint arXiv:2305.04930}, 2023.

\bibitem{xiao2023star}
H.~Xiao, X.~Hu, T.-X. Zheng, and K.-K. Wong, ``{STAR-RIS Assisted Covert
  Communications in NOMA Systems},'' \emph{arXiv preprint arXiv:2306.07105},
  2023.

\bibitem{forouzesh2020joint}
M.~Forouzesh, P.~Azmi, A.~Kuhestani, and P.~L. Yeoh, ``Joint
  information-theoretic secrecy and covert communication in the presence of an
  untrusted user and warden,'' \emph{IEEE Internet Things J.}, vol.~8, no.~9,
  pp. 7170--7181, 2020.

\bibitem{akdeniz2014millimeter}
M.~R. Akdeniz, Y.~Liu, M.~K. Samimi, S.~Sun, S.~Rangan, T.~S. Rappaport, and
  E.~Erkip, ``Millimeter wave channel modeling and cellular capacity
  evaluation,'' \emph{IEEE J. Sel. Areas Commun.}, vol.~32, no.~6, pp.
  1164--1179, 2014.

\bibitem{wu21}
C.~Wu, C.~You, Y.~Liu, X.~Gu, and Y.~Cai, ``{Channel estimation for
  {STAR-RIS}-aided wireless communication},'' \emph{IEEE Commun. Lett.},
  vol.~26, no.~3, pp. 652--656, 2021.

\bibitem{Wang21}
C.~Wang, Z.~Li, J.~Shi, and D.~W.~K. Ng, ``{Intelligent Reflecting
  Surface-Assisted Multi-Antenna Covert Communications: Joint Active and
  Passive Beamforming Optimization},'' \emph{IEEE Trans. Commun.}, vol.~69,
  no.~6, pp. 3984--4000, 2021.

\bibitem{luo2010semidefinite}
Z.-Q. Luo, W.-K. Ma, A.~M.-C. So, Y.~Ye, and S.~Zhang, ``Semidefinite
  relaxation of quadratic optimization problems,'' \emph{IEEE Signal Process.
  Mag.}, vol.~27, no.~3, pp. 20--34, 2010.

\bibitem{grant14cvx}
M.~Grant and S.~Boyd, ``{CVX: Matlab software for disciplined convex
  programming, version 2.1},'' 2014.

\bibitem{boyd04}
S.~Boyd, S.~P. Boyd, and L.~Vandenberghe, \emph{Convex optimization}.\hskip 1em
  plus 0.5em minus 0.4em\relax Cambridge Univ. Press, 2004.

\bibitem{feng2022joint}
C.~Feng, W.~Shen, J.~An, and L.~Hanzo, ``Joint hybrid and passive
  {RIS}-assisted beamforming for mmwave {MIMO} systems relying on dynamically
  configured subarrays,'' \emph{IEEE Internet Things J.}, vol.~9, no.~15, pp.
  13\,913--13\,926, 2022.

\bibitem{zhang2017matrix}
X.-D. Zhang, \emph{Matrix analysis and applications}.\hskip 1em plus 0.5em
  minus 0.4em\relax Cambridge Univ. Press, 2017.

\bibitem{evans00}
J.~Evans and D.~N.~C. Tse, ``Large system performance of linear multiuser
  receivers in multipath fading channels,'' \emph{IEEE Transa. Inf. Theory},
  vol.~46, no.~6, pp. 2059--2078, 2000.

\end{thebibliography}

\end{document}